\documentclass[twocolumn,pra,longbibliography,superscriptaddress]{revtex4-1}
\usepackage[pdftex]{graphicx}
\usepackage{amsmath,amsfonts,amsbsy,amssymb,amsthm}
\usepackage{mathtools,xfrac}

\usepackage[percent]{overpic}

\usepackage[colorlinks]{hyperref}
\usepackage{siunitx}
\usepackage[export]{adjustbox}

\renewcommand{\Re}{\text{Re}}
\renewcommand{\Im}{\text{Im}}

\newcommand{\tr}[1]{\text{tr}\left\{ #1 \right\}}
\newcommand{\Tr}[1]{\text{Tr}\left\{ #1 \right\}}

\usepackage[caption=false]{subfig}

\newcommand{\TJU}{College of Precision Instrument and Opto-Electronics Engineering, Key Laboratory of
Opto-Electronics Information Technology, Ministry of Education,
Tianjin University, Tianjin 300072, China.\looseness=-1}
\newcommand{\ANU}{Centre for Quantum Computation and Communication Technology, Department of Quantum Science,
Research School of Physics and Engineering, Australian National University, Canberra ACT 2601, Australia.\looseness=-1}
\newcommand{\IU}{Department of Physics, Indiana University-Purdue University Indianapolis, Indianapolis, IN 46202, USA.\looseness=-1}
\newcommand{\NTU}{School of Physical and Mathematical Sciences, Nanyang Technological University, Singapore 639673.\looseness=-1}

\begin{document}

\title{Accessible precisions for estimating two conjugate parameters using Gaussian probes}
\author{Syed M. Assad}
\email{cqtsma@gmail.com}
\affiliation{\TJU}
\affiliation{\NTU}
\affiliation{\ANU}
\author{Jiamin Li}
\affiliation{\TJU}
\author{Yuhong Liu}
\affiliation{\TJU}
\author{Ningbo Zhao}
\affiliation{\TJU}
\author{Wen Zhao}
\affiliation{\TJU}
\author{Ping Koy Lam}
\affiliation{\ANU}
\author{Z. Y. Ou}
\email{zou@iupui.edu}
\affiliation{\TJU}
\affiliation{\IU}
\author{Xiaoying Li}
\email{xiaoyingli@tju.edu.cn}
\affiliation{\TJU}

\begin{abstract}
We analyse the precision limits for simultaneous estimation
of a pair of conjugate parameters in a displacement channel using
Gaussian probes. Having a set of squeezed states as an
initial resource, we compute the Holevo Cramér-Rao bound to
investigate the best achievable estimation precisions if only
passive linear operations are allowed to be
performed on the resource prior to probing the channel. The analysis reveals the
optimal measurement scheme and allows us to quantify the best
precision for one parameter when the precision of the second
conjugate parameter is fixed. To estimate the conjugate parameter
pair with equal precision, our analysis shows that the optimal
probe is obtained by combining two squeezed states with
orthogonal squeezing quadratures on a 50:50 beam splitter. If
different importance are attached to each parameter, then the
optimal mixing ratio is no longer 50:50. Instead it follows a
simple function of the available squeezing and the relative
importance between the two parameters.
\end{abstract}
\date{March 16, 2020}
\maketitle

\section{Introduction}
How precise can we make a set of physical measurements? This is a
fundamental question that has driven much of the progress in science
and technology. Improving the precisions and understanding limitations
to measurements have often led to revolutionary discoveries or new
insights in science. After overcoming technical sources of noise, the
presence of quantum noise imposes a limit to the ultimate measurement
precision. Due to the presence of quantum fluctuations, estimation
precision using classical probe fields is limited to the {\em standard
  quantum limit} for optical measurements. In order to surpass this
limit, quantum resource such as squeezed
states~\cite{Caves1981,Xiao1987,Grangier1987} or entangled
states~\cite{Dariano2001,Fujiwara2001,Fischer2001,Sasaki2002,
  Fujiwara2003,Ballester2004,Giovannetti2004,Genoni2013,Rigovacca2017,
  Bradshaw2017,Bradshaw2018,Liu2018,Li2018,Gupta2018} are required. A
notable example is the use of quadrature squeezed states of light to
enhance the detection of gravitational wave~\cite{aasi2013,grote2013}.
Another concept in quantum mechanics that distinguishes it from
classical mechanics is that of non-commuting observables. This imposes
a limitation for simultaneously estimating multiple parameters encoded
in non-commuting observables.

In this work, we consider the problem of estimating two independent
parameters $\theta = (\theta_x,\theta_y)$, encoded in two conjugate
quadratures $X$ and $Y$ of a displacement channel
$D(\theta)=\exp\left(\frac{i\theta_y}{2} X - \frac{i \theta_x}{2}
  Y\right)$. This channel induces a displacement of $\theta_x$ on the
amplitude quadrature $X$ and $\theta_y$ on the phase quadrature $Y$ of
a single-mode optical field with $[ X, Y]=2i$. This problem has
attracted a lot of attention since the early days of quantum
mechanics~\cite{Arthurs1965,Yuen1982,Arthurs1988} and continue to do
so~\cite{Duivenvoorden2017,Genoni2013,Bradshaw2018}. For example, if a
single-mode probe is used to sense the displacement, the work by
Arthurs and Kelly showed that the estimation mean squared errors $v_x$
and $v_y$ are bounded by $ v_x v_y \geq
4$~\cite{Arthurs1965}. However, it was theoretically
shown~\cite{Braunstein2000,Zhang2000,Li2018} and experimentally
demonstrated~\cite{Li2002,Steinlechner2013,Liu2018} that by utilising quantum
entanglement between two systems---for example through the quantum
dense coding scheme---it is possible to circumvent this limit and
estimate both parameters with accuracies beyond the standard quantum
limit.

More recently, the pioneering works by Holevo and Helstrom on quantum
estimation
theory~\cite{Helstrom1967,Helstrom1969,Holevo1976,*Holevo2011} have
been used to study this
problem~\cite{Genoni2013,Gao2014,Bradshaw2017,Bradshaw2018}. Once the
probe state is specified, the quantum Fisher information determines a
bound on the estimation precision thorough the quantum Cram\'er-Rao
bound (CRB), which holds for every possible measurement
strategy. There are many variants of the quantum CRB---the two most
popular being the symmetric logarithmic derivative
(SLD)~\cite{Helstrom1967,Helstrom1969,Braunstein1994,Fujiwara1995} and
the right logarithmic derivative
(RLD)~\cite{Yuen1973,Belavkin1976,Fujiwara1994,Fujiwara1994a,Fujiwara1995,Fujiwara1999}
as these yield direct bounds for the sum of the mean squared
error. These have been widely used since they are relatively easy to
compute~\cite{Paris2009,Petz2011}. For single-parameter estimation,
the SLD-CRB offers an asymptotically tight bound on the
precision~\cite{Barndorff-Nielsen2000}. However for multi-parameter
estimation, neither the SLD-CRB nor the RLD-CRB is necessarily
tight~\cite{Szczykulska2016,Suzuki2019}. Hence even though the probe
might offer a large quantum Fisher information, their CRB might not be
{\em achievable}, which means that the actual achievable precisions
are not known.

Here, we solve this problem by using the Holevo Cram\'er-Rao bound to
compute the actual asymptotically achievable
precision~\cite{Holevo1976,*Holevo2011,Nagaoka2005,Hayashi2006,Yamagata2013}. Knowing
the achievable precision for a specific probe allows us to compare
metrological performances between two different probes. We can then
use this formalism to answer the question: Given a fixed quantum
resource such as squeezing, how do we use it to optimally sense the
channel? The resource states that we consider will be one-mode and
two-mode Gaussian states, which we are allowed to freely mix or rotate
before sending one mode to probe the channel. In doing so, we derive
ultimate bounds on simultaneous parameter estimation which goes beyond
existing restrictions imposed by the SLD or RLD-CRB. These bounds
quantify a {\em resource apportioning principle}---the resource can be
allocated to gain either a precise estimate of $\theta_x$ or
$\theta_y$ but not both together~\cite{Li2018,Liu2019}.

The paper is organised as follows. We start with a summary of the
general framework for two-parameter estimation in
section~\ref{sec:framework}. Next we apply this framework to derive
from the Fisher information precision limits for a single mode probe
in section~\ref{sec:single_mode}. We then generalise this result to
two-mode probes in section~\ref{sec:two_mode}. We show that at least
\SI{6}{dB} of squeezing is necessary to surpass the standard quantum
limit. We also elucidate our results with two examples: the first with
a single squeezed state and the second with two squeezed states with
equal amount of squeezing. Finally, we end with some discussions in
section~\ref{sec:discussions}.

\begin{center}
\begin{figure*}[t]
  \subfloat{\label{fig:1a}
    \includegraphics[width=0.4\columnwidth,valign=m]{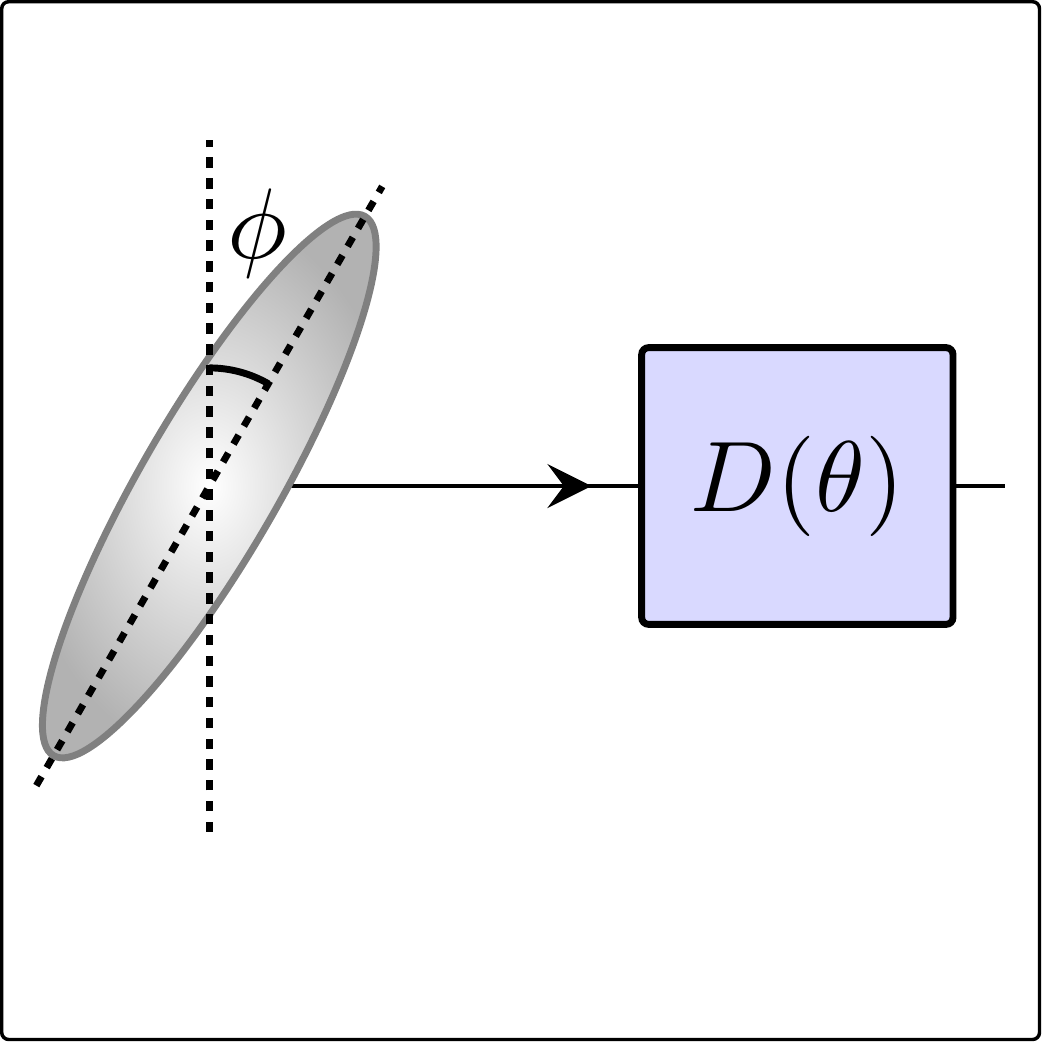}
    \vphantom{  \includegraphics[width=0.5\columnwidth,valign=m]{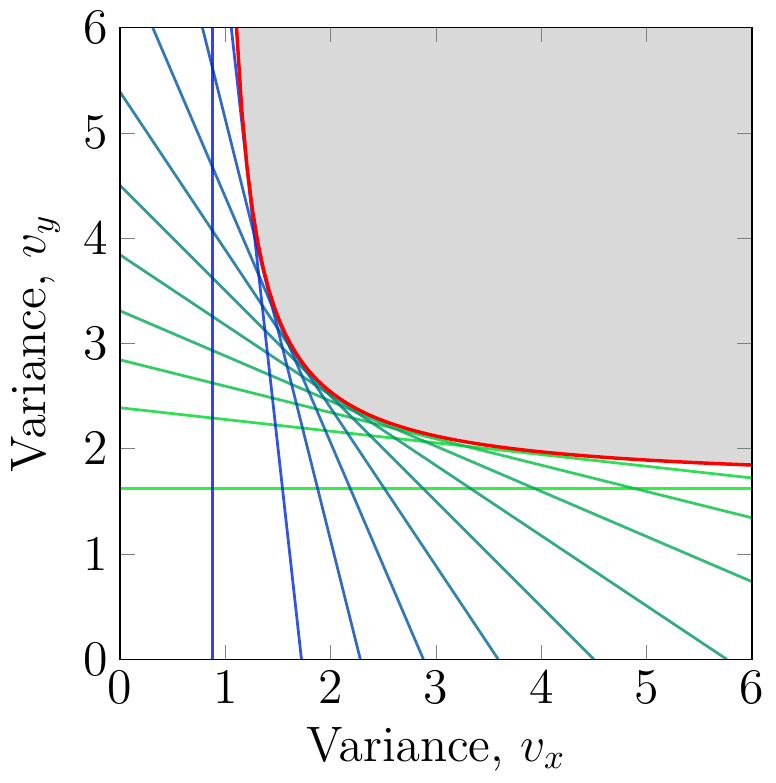}}}%
  \put(-19,40){(a)}
\subfloat{\label{fig:1b}
  \includegraphics[width=0.5\columnwidth,valign=m]{./figures/fig1b-figure0.pdf}}
  \put(-19,45){(b)}
  \subfloat{\label{fig:1c}
    \includegraphics[width=0.5\columnwidth,valign=m]{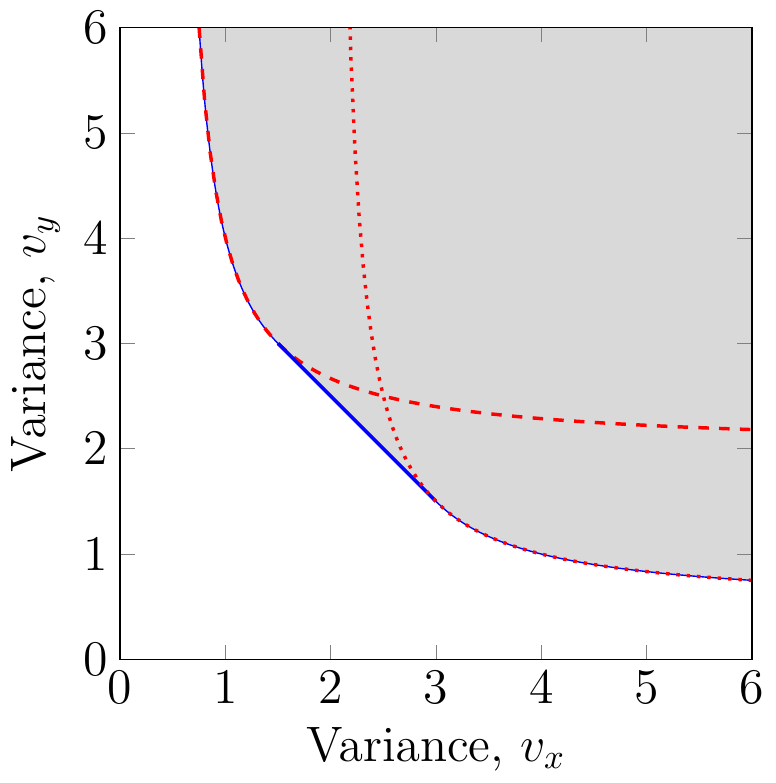}}
  \put(-19,45){(c)}
\subfloat{\label{fig:1d}
    \includegraphics[width=0.51\columnwidth,valign=m]{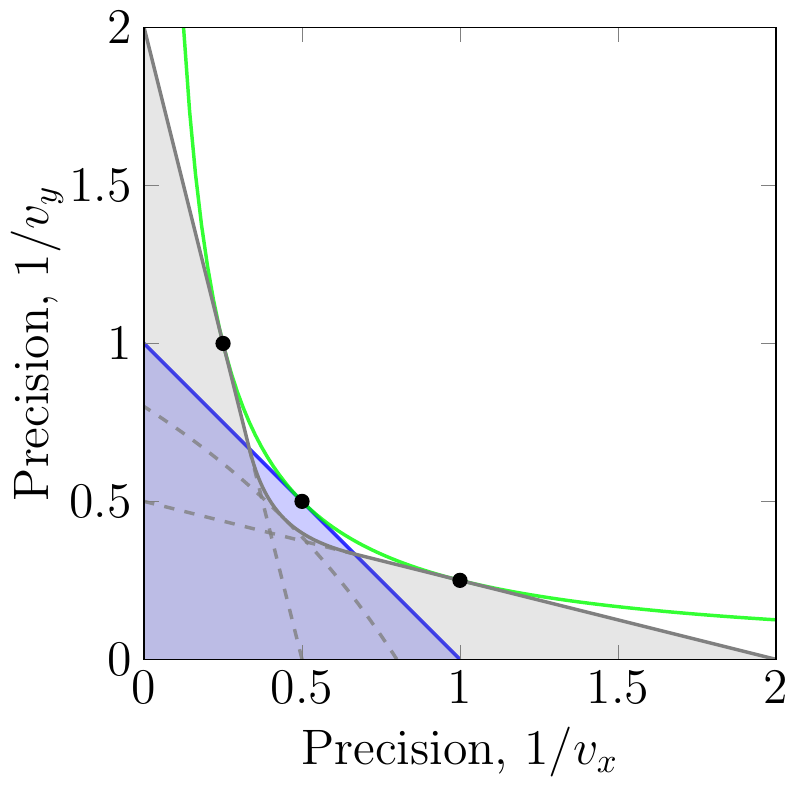}}
  \put(-19,45){(d)}
  \caption{ (a) A squeezed state is used to sense the parameter
    $\theta$ of a displacement channel. (b) With 3 dB of squeezing,
    and for a fixed squeezing angle $\phi=\pi/6$, each of the straight
    line is the Holevo-CRB~(\ref{sinH}) with a different value of
    $w_x/w_y$. The shaded area shows the accessible variance for
    simultaneously estimating $\theta_x$ and $\theta_y$. (c) The two
    red dashed and dotted lines can be achieved by an $X$ and $P$
    squeezed state with $\phi=0$ and $\phi=\pi/2$ respectively. The
    blue line requires an intermediate squeezing angle. The shaded
    area are all the accessible regions for a single mode squeezed
    state.  (d) This shows the same region as (c) but as a function of
    the precision.  With a \SI{3}{dB} squeezed state, we can reach the
    grey areas. More squeezing can give a high precision for one
    parameter but at the expense of a lower precision for the other. The
    product of the precisions will never exceed $1/4$ regardless of the
    squeezing level. This is shown as the green line.  The three grey
    dashed lines plot Eq.~(\ref{snrB1}) when the squeezing angles are
    fixed at $\phi=0$, $\pi/4$ and $\pi/2$. The vacuum probe can only
    access the blue region.}
\end{figure*}
\end{center}

\section{General framework}
\label{sec:framework}
Let us begin with a brief review of the two-parameter estimation
problem and the Holevo Cram\'er-Rao bound.  To estimate the parameters
$\theta$, the state $\rho_0$ is sent through the displacement channel
$D(\theta)$ as a probe. After the interaction, the state becomes
$\rho_\theta= D(\theta) \rho_0 D(\theta)^\dagger$ which now contains
information about the two parameters of interest. Next, we
perform some measurement scheme and use an estimation strategy which
leads to two unbiased estimators $\hat{\theta}_x$ and
$\hat{\theta}_y$. We quantify the performance of these estimators,
through the mean squared errors
\begin{align}
  v_x \coloneqq \mathbb{E}\left[
  (\hat{\theta}_x-\theta_x)^2\right]\;\text{ and }\;
    v_y \coloneqq \mathbb{E}\left[
  (\hat{\theta}_y-\theta_y)^2\right]\;.
\end{align}
When restricted to classical probes, due to quantum noise we have
$v_x\geq 1$ and $v_y\geq 1$ which is known as the standard quantum
limit. The aim of this work is to find out what are the possible values
that $v_x$ and $v_y$ can take simultaneously. To quantify the
performance for estimating both $\theta_x$ and $\theta_y$
simultaneously, we use the weighted sum of the mean squared error:
$w_x v_x+w_y v_y$ as a figure of merit where $w_x$ and $w_y$ are
positive weights that quantify the importance we attach to parameters
$\theta_x$ and $\theta_y$ respectively.  We want to find an estimation
strategy that minimises this quantity.

The Holevo-CRB sets an
asymptotically attainable bound on the weighted sum of the mean
squared error~\cite{Holevo1976,*Holevo2011}
\begin{align}
  \label{HCR}
w_x v_x + w_y v_y \geq  f_\text{HCR}:=\min_{\mathcal{X}} h_\theta[\mathcal{X}]\;,
\end{align}
where $\mathcal{X}= \left\{ \mathcal{X}_x, \mathcal{X}_y \right\}$ are
Hermitian operators that satisfy the locally unbiased conditions
  \begin{align}
   \label{con}
\left. \tr{\rho_\theta \mathcal{X}_j}\right\rvert_{\theta=0}= 0\;
    \text{ and }\;
\left. \tr{\frac{\partial \rho_\theta}{\partial \theta_j}\mathcal{X}_k }\right\rvert_{\theta=0} =\delta_{jk}    \;,
\end{align}
for $j,k \in \{x,y\}$ and $h_\theta$ is the function
\begin{align}
  h_\theta[\mathcal{X}]\coloneqq \Tr{W\, \Re Z_\theta[\mathcal{X}]}
  +\left\lVert \sqrt{W}\,\Im Z_\theta[\mathcal{X}]\sqrt{W} \right\rVert_1\;.
\end{align}
Here $Z$ is the 2-by-2 matrix
$ Z_{jk} \coloneqq \tr{\rho_\theta \mathcal{X}_j \mathcal{X}_k}$ and
$W$ is a diagonal matrix with entries $w_x$ and $w_y$. The
bound depends on the state $\rho_\theta$ only; it does not need for us
to specify any measurement. For quadrature displacements with Gaussian
probes, the bound involves minimisation of a convex function over a
convex domain. This is an instance of convex optimisation problem
which can be calculated efficiently by numerical
methods~\cite{Bradshaw2018}. Furthermore, the optimisation also
reveals an explicit measurement scheme that saturates the bound. For
Gaussian probes, the optimal measurement will always be an individual
Gaussian measurement.

\section{Precision bounds for single-mode probe}
\label{sec:single_mode}
We now apply the formalism to a pure single-mode amplitude squeezed
state probe with quadrature variance $e^{-2r}$ and rotated by an angle
$\phi$ as shown in Fig.~\ref{fig:1a}. As previously stated, the
Holevo-CRB only depends on the probe and how it varies with the
parameters. In the single mode case, constraints~(\ref{con}) fully
determines $f_\text{HCR}$.  There is no free parameter in the
optimisation and as a result, Holevo-CRB~(\ref{HCR}) becomes
\begin{align}
  \label{sinH}
  w_x v_x + w_y v_y \geq w_x v_a + w_y v_b +2\sqrt{w_x w_y}\;,
\end{align}
where
\begin{align}
  \label{vx1}
v_a  &\coloneqq e^{-2r}\cos^2 \phi + e^{2r}\sin^2 \phi \;,\\
v_b  &\coloneqq e^{-2r}\sin^2 \phi + e^{2r}\cos^2 \phi \;,
\end{align}
are the projected variances on the $X$ and $Y$ quadratures. For every
choice of $w_x/w_y$, Eq.~(\ref{sinH}) defines a straight line in the
$v_x$--$v_y$ plane and gives a different bound on that plane. Some of
these bounds are plotted in Fig.~\ref{fig:1b} for $e^{-2r}=1/2$ and
$\phi=\pi/6$. For example, to estimate both $\theta_x$ and $\theta_y$ with equal
precision, setting $w_x=w_y=1$ gives the best estimation strategy
with $v_x+v_y=2(1+\cosh 2r)$ independent of $\phi$. This gets worse with more
squeezing. However, if we are only concerned with estimating $\theta_x$,
setting $w_y=0$ results in $v_x=v_a$. By eliminating $w_x$ and
$w_y$ from Eq.~(\ref{sinH}), we can collect all these bounds into one
stricter bound
\begin{align}
  \label{snrB1}
   (v_y - v_b) (v_x- v_a) \geq 1\;
\end{align}
which holds for every $\phi$. This is plotted in Fig.~\ref{fig:1c} for
a few vales of $\phi$. Every pair of $(v_x, v_y)$ that satisfies
Eq.~(\ref{snrB1}) can be achieved by a specific measurement
strategy. The same relation is plotted in Fig.~\ref{fig:1d} as a
function of the precisions $ 1/v_x$ and $ 1/v_y$. This relation
quantifies the resource apportioning principle---given a fixed amount
of squeezing, there is only so much improvement in the precision to be
had. The resource can be used to gain a precise estimate of
$\theta_x$, but this comes at the expense of an imprecise estimate of
$\theta_y$.

When $\phi=0$, relation~(\ref{snrB1}) can be written concisely as
a bound on the weighted sum of the precisions
\begin{align}
  \label{sinB}
  \frac{ e^{-2r}}{v_x}+ \frac{ e^{2r}}{v_y} \leq 1\;.
\end{align}
By using the arithmetic-geometric mean inequality, an immediate
corollary of the result is the Arthurs and Kelly relation
$v_x v_y \geq 4$ which holds for all
$r$~\cite{Arthurs1965,Li2018}. This reflects the Heisenberg
uncertainty relation imposed on a single mode system. Every value of
squeezing can saturate this inequality at one value of $v_x$ and $v_y$
as seen in Fig~\ref{fig:1d}. As we shall show next, this restriction
can be somewhat relaxed using two mode states, but the sum of the
precisions are still constrained by the total available resource.

\begin{center}
\begin{figure*}[t]
  \subfloat{\label{fig:2a}
    \includegraphics[width=0.45\columnwidth,valign=m]{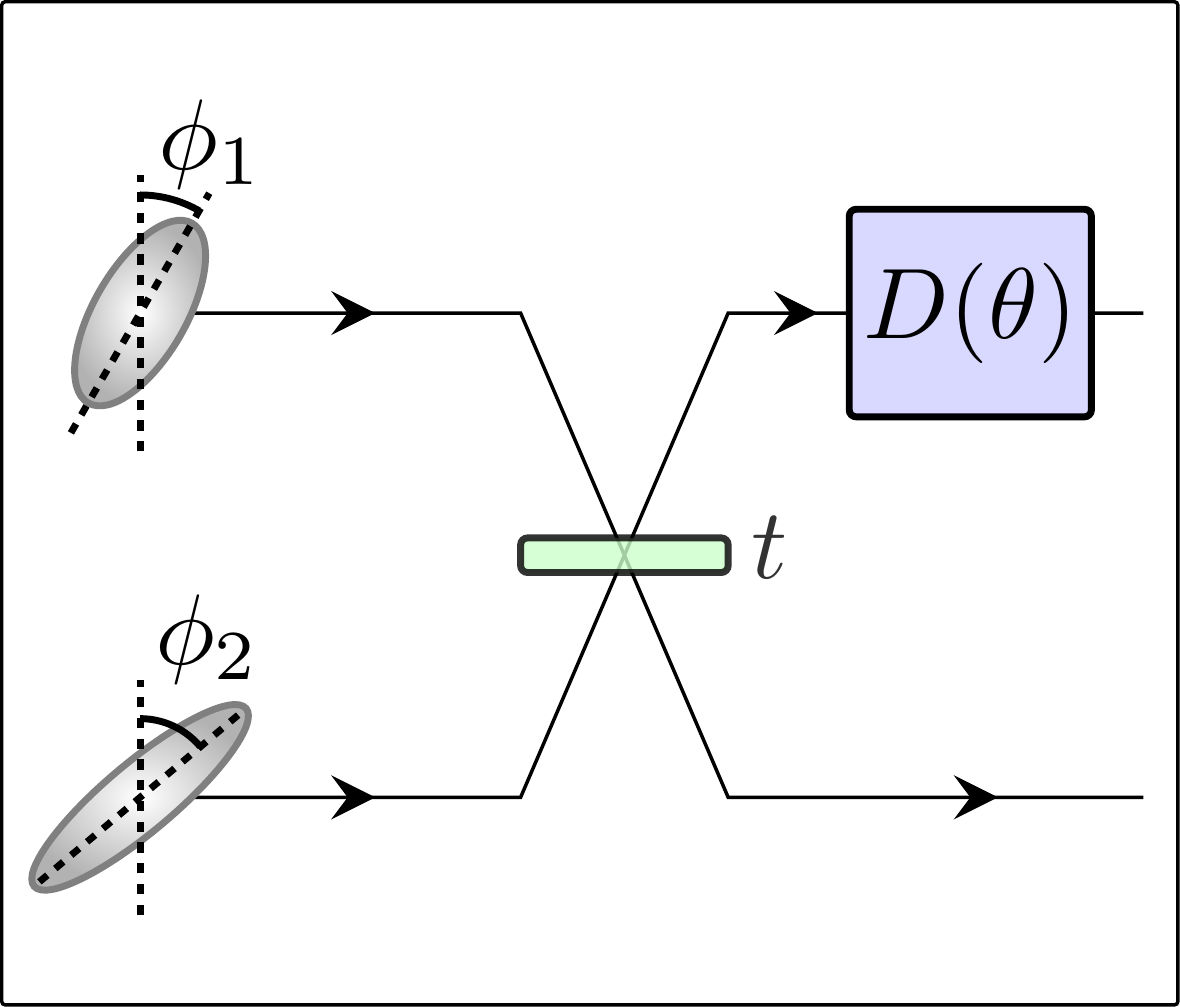}}
  \put(-17,40){(a)}
\subfloat{\label{fig:2b}
\includegraphics[width=0.5\columnwidth,valign=m]{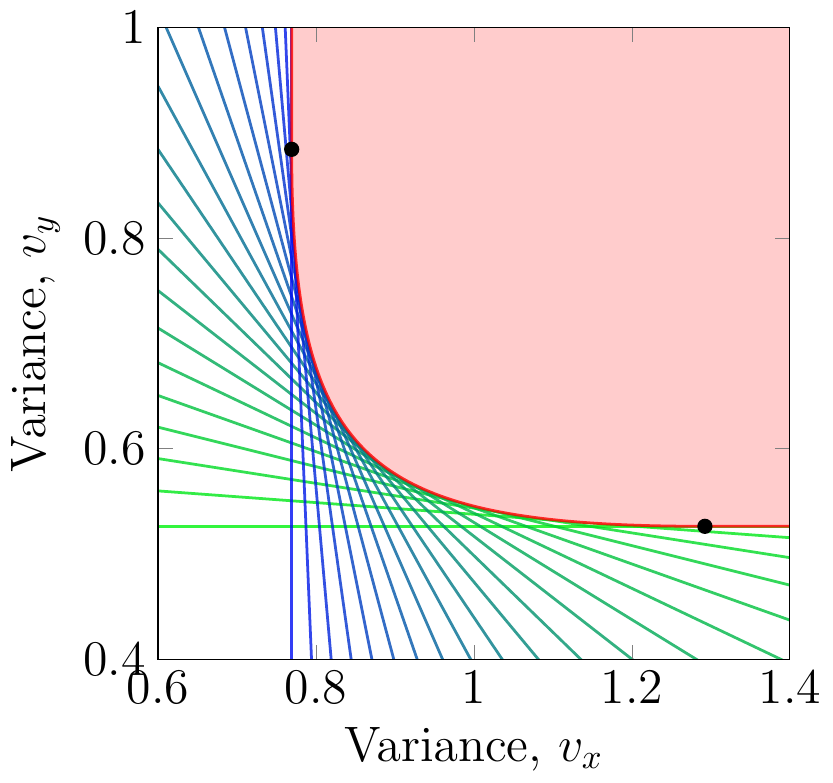}}
  \put(-20,45){(b)}
  \subfloat{\label{fig:2c}
    \includegraphics[width=0.5\columnwidth,valign=m]{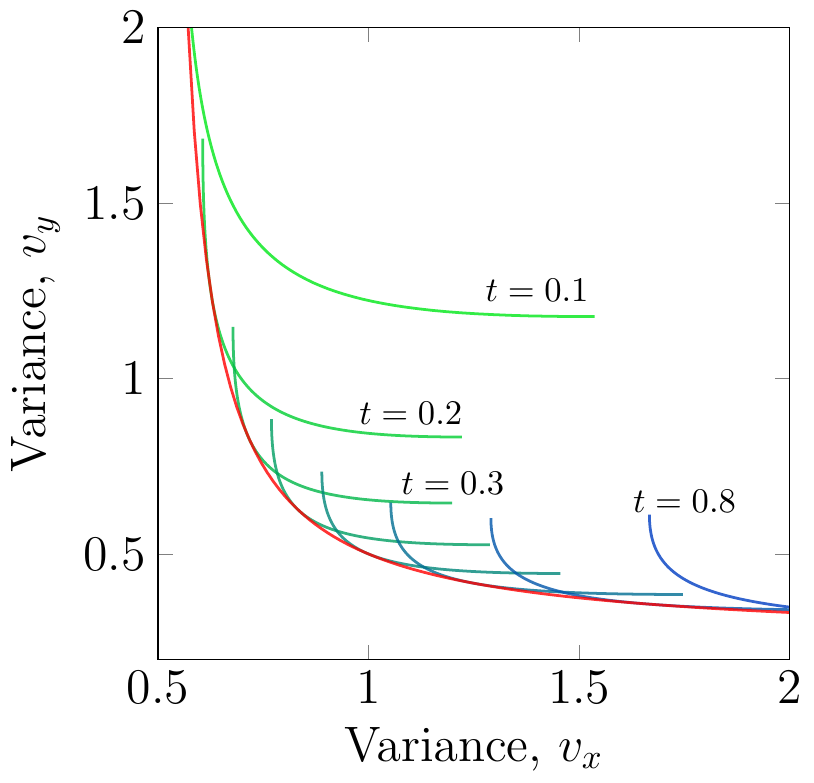}}
  \put(-17,45){(c)}
  \subfloat{\label{fig:2d}
    \includegraphics[width=0.5\columnwidth,valign=m]{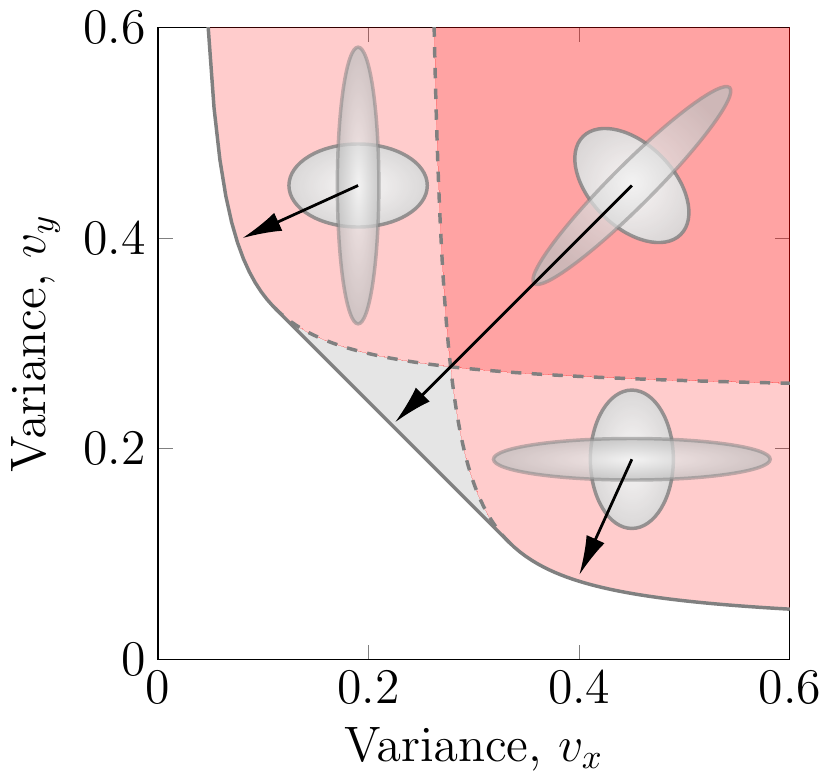}}
  \put(-19,45){(d)}
  \caption{(a) Two squeezed states are used to sense the displacement
    $\theta$. (b) The Holevo-CR bound for a two-mode probe with
    $r_1=0.35$, $r_2=0.69$, $t=0.4$, $\phi_1=0$ and
    $\phi_2=\pi/2$. Each straight line correspond to a bound with
    different values of $w_x/w_y$. The pink region shows all the
    accessible values of $v_x$ and $v_y$.  (c) Each bluish-green curve
    gives the accessible boundary for the same probe as (b) except for
    the value of $t$ which varies from $0.1$ to $0.8$ in steps of
    $0.1$. The red curve is the envelope of all the blueish-green
    curve. (d) The shaded areas show the relation~(\ref{eq:3a})
    having two squeezed probes with \SI{6}{dB} and \SI{15.6}{dB} of
    squeezing. The variance for estimating both parameters can be simultaneously
    smaller than 1. The two grey dashed lines are limits when the
    probe is fixed with $\phi_1=0$ and $\pi/2$ given by
    Eq.~(\ref{eq:SNR0}). }
\end{figure*}
\end{center}
\section{Precision bounds for two-mode probe}
\label{sec:two_mode}
We now consider a two-mode system where we have access to two
amplitude-squeezed states with quadrature variances $e^{-2r_1}$ and
$e^{-2r_2}$. Furthermore we are allowed to rotate them by $\phi_1$ and
$\phi_2$, and mix the two through a beam-splitter of transmissivity
$t$ before sending one mode through the displacement channel as shown
in Fig.~\ref{fig:2a}. In this case, $f_\text{HCR}$ does not have
a simple form; its computation involves finding the root of a quartic
function. Despite this, the collection of all the bounds lead to a
final expression that is surprisingly simple and intuitive. This is
our main result: Given two pure squeezed states with variances
$e^{-2r_1}$ and $e^{-2r_2}$ as a resource where $0\leq r_1 \leq r_2$,
and allowing for rotation and mixing operations, the measurement
sensitivity is limited by
\begin{align}
  \label{eq:3a}
  v_y\geq  v_y^*
        &=\left\{
          \begin{alignedat}{2}
          \frac{v_x e^{-2r_1}}{v_x-e^{-2r_2}} &\text{ if }
          &e^{-2r_2} \leq v_x<v_c\\
          \left(e^{-r_1}+e^{-r_2} \right)^2 -v_x &\text{ if }
          &v_c \leq v_x< v_d\\
          \frac{v_x e^{-2r_2}}{v_x-e^{-2r_1}}  &\text{ if }  & v_d \leq v_x 
          \end{alignedat}
\right.
\end{align}
where $v_c \coloneqq e^{-2r_2}+e^{-r_1-r_2}$ and
$v_d \coloneqq e^{-2r_1}+e^{-r_1-r_2}$. The full derivation requires a
lengthy but straightforward minimization and is relegated to the
supplementary section. It involves finding the optimal values of
$\phi_1$, $\phi_2$ and $t$ for every pair of $w_x$ and $w_y$.  We
outline the main steps in the derivations here. Firstly, for a fixed
value of $w_x$ and $w_y$ and $t$, we can numerically compute the
Holevo-CRB for each pair of $\phi_1$ and $\phi_2$. We find that the
optimal setting for $\phi_2$ is when $\phi_2=\phi_1+\pi/2$, making the
two squeezed states as different as
possible~\cite{Olivares2011}. Secondly, for a fixed $\phi_1$ and $t$,
each pair of $w_x$ and $w_y$ gives a bound which correspond to one of
the straight lines plotted in Fig~\ref{fig:2b}. The collection of all
these bounds give the accessible region for this probe configuration. Thirdly,
we vary $t$ to find the accessible region for a fixed $\phi_1$ as
shown in Fig~\ref{fig:2c}. Finally the optimal value of $\phi_1$ is
determined to arrive at the final result~(\ref{eq:3a}).

The region described by~(\ref{eq:3a}) is plotted in
Fig.~\ref{fig:2d}. Every pair of $(v_x,v_y)$
that satisfies relation~(\ref{eq:3a}) can be attained by a dual
homodyne measurement. An immediate corollary of this is the relation
$v_x v_y \geq 4 e^{-2r_1} e^{-2r_2}$~\cite{Steinlechner2013}. In order to surpass the standard quantum
limit for both parameters, we require $e^{-2r_1} e^{-2r_2} < {1}/{4} $. In other
words, the sum of the squeezed variances of the resource has to be
greater than approximately \SI{6}{\dB}.

As mentioned in the outline of the derivations, not all regions
in~(\ref{eq:3a}) can be reached using the same probe. Different region
requires the resource to be used differently. For $w_x<w_y$, the best
way to use the available resource is to set $\phi_1=0$ and
$\phi_2=\pi/2$ and mix them on a beam-splitter with transmissivity
\begin{align}
  \label{tstar}
  t=\frac{e^{r_1}}{e^{r_1}+e^{r_2}\sqrt{w_x/w_y}}\;.
\end{align}
This gives the optimal variances
\begin{align}
  v_x &=e^{-2r_1}+e^{-(r_1+r_2)}\sqrt{w_y/w_x}\;,\\
  v_y &=e^{-2r_2}+e^{-(r_1+r_2)}\sqrt{w_x/w_y}\;,
\end{align}
or in terms of $t$,
\begin{align}
  v_x =\frac{e^{-2r_1}}{1-t}\;\text{ and }\;
  v_y =\frac{e^{-2r_2}}{t}
\end{align}
for $t>\frac{e^{r_1}}{e^{r_1}+e^{r_2}}$. After eliminating $t$, we
arrive at a bound on the precisions
\begin{align}
\label{eq:SNR0}
  \frac{e^{-2r_1}}{v_x} + \frac{ e^{-2r_2}}{v_y} \leq 1\;.
\end{align}
For $w_y<w_x$, we just need to
swap the roles of $x$ and $y$ by setting $\phi_1=\pi/2$ and
$\phi_2=0$. Equations~(\ref{tstar})--(\ref{eq:SNR0}) still hold with
all $x$ and $y$ swapped. When $w_x=w_y$, there is
a family of estimation strategy that all give the same sum of
variances $v_x +v_y=(e^{-r_1}+e^{-r_2})^2$ but different values for
each individual variances. This can be accessed by varying $\phi_1$
from $0$ to $\pi/2$ with $\phi_2=\phi_1+\pi/2$ and keeping $t$ as
Eq.~(\ref{tstar}) which gives
\begin{align}
\label{eq:wxewy}
  \left.  \begin{array}{c}v_x\\v_y\end{array} \right\}
  =\frac{1}{2}\left(e^{-r_1}+e^{-r_2}\right)^2\pm\frac{\cos 2\phi_1}{2}
  \left(e^{-2r_1}-e^{-2r_2} \right)\;.
\end{align}

In the following, we illustrate these results with two
examples. In these example, we present the optimal probe and
measurement strategy that saturates the estimation
precisions~(\ref{eq:3a}).

\subsection{Example 1: One squeezed state and one vacuum}

\begin{figure}[t]
    \includegraphics[width=0.7\columnwidth]{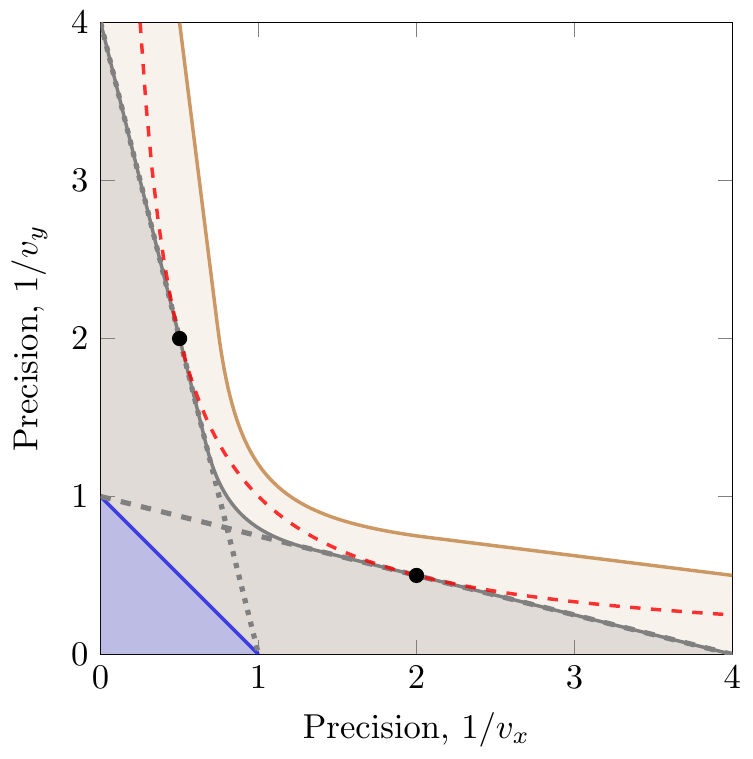}
    \put(-117,95){    \includegraphics[scale=0.3]{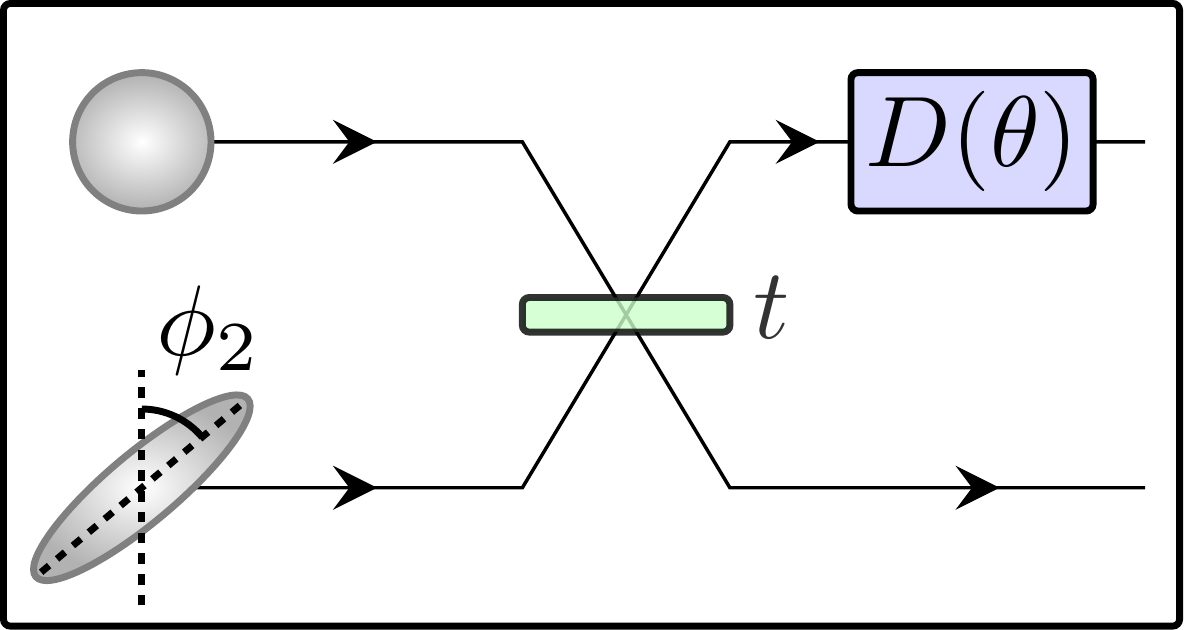}}
    \caption{\label{fig:3} In order to surpass the standard quantum
      limit, $v_x v_y =1$ (red dashed line), we require access to an
      additional ancillary mode. The accessible region for a squeezed
      state with \SI{6}{dB} of squeezing is shown as the grey shaded
      region. It can just reach the standard quantum limit at the two
      black dots. The dashed and dotted grey lines plot Eqs.~(\ref{eq:sxy0})
      and~(\ref{eq:sxy1}) which can be accessed by setting
      $\phi_2=\pi/2$ and $\phi_2=0$ respectively. With \SI{9}{dB} of
      squeezing, we can clearly surpass this limit (brown
      region). These bounds are given by Eq.~(\ref{eq:3a}).}
\end{figure}
\begin{figure}[t]
    \centering
\includegraphics[width=0.8\columnwidth]{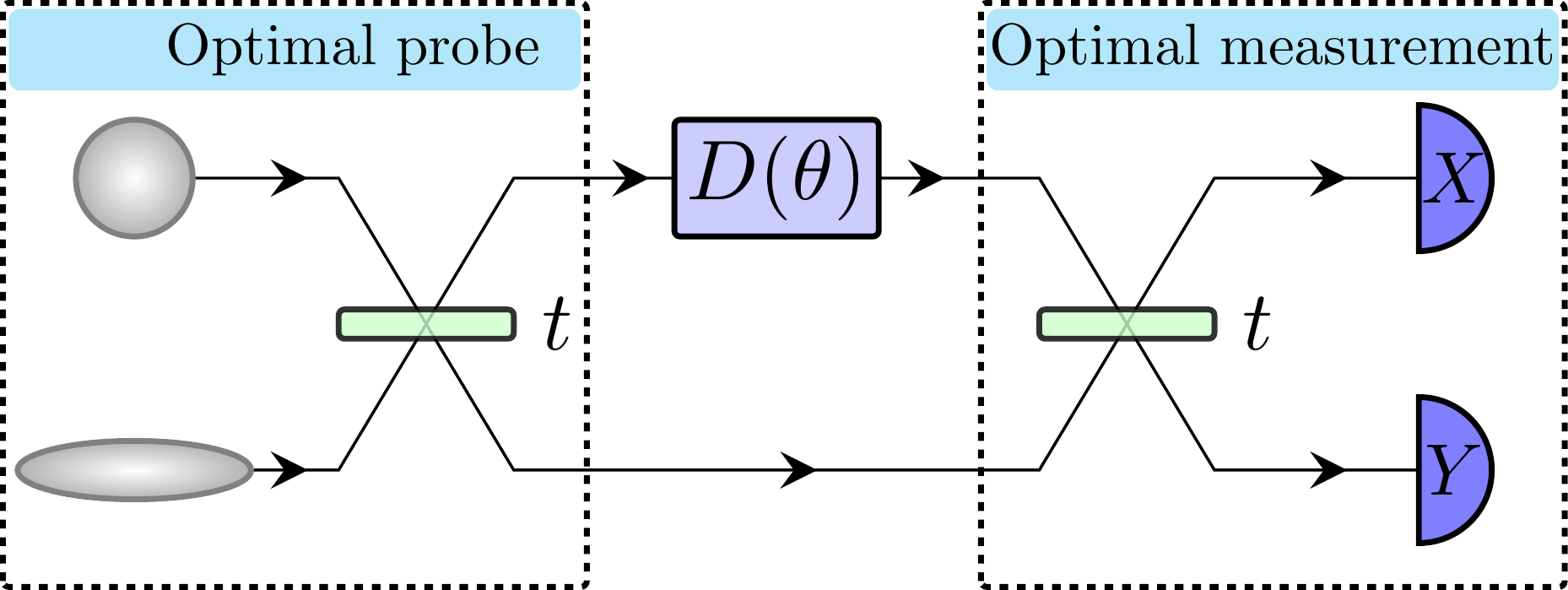}
\caption{
  \label{fig:4}With one squeezed state and for $w_x<w_y$, the optimal
  probe configuration is to prepare a $Y$-squeezed and split it on a
  beam-splitter with $t=\frac{1}{1+\sqrt{w_x/w_y}}$. The optimal
  measurement is to disentangle the two modes on a second
  beam-splitter and perform $X$ and $Y$ quadrature measurements on the
  two outputs which gives the variances in~(\ref{eq:1}).}
\end{figure}
In our first example, we consider the case of one squeezed state
and one vacuum state ($r_1=0$) as shown in Fig.~\ref{fig:3} inset.

For $w_x < w_y$, the optimal use of the probe is to set $\phi_2=\pi/2$ and
the optimal measurement setup is shown in Fig.~\ref{fig:4}. The two
quadrature measurements give independent estimates of $\theta_x$ and
$\theta_y$ with variances
\begin{align}
  \label{eq:1}
  v_x=\frac{1}{1-t}\;\text{ and }\; v_y=\frac{e^{-2r_2}}{t}\;.
\end{align}
For $\frac{1}{1+e^{r_2}} \leq t \leq 1$, this pair of variances is
optimal.  Eliminating $t$, we can improve on the single mode
precision relation~(\ref{sinB}) with
\begin{align}
  \label{eq:sxy0}
  \frac{1}{v_x}  + \frac{e^{-2r_2}}{v_y}    \leq 1\;
\end{align}
which is plotted as the dashed grey line in Fig.~\ref{fig:3} for $e^{-2r_2}=1/4$.
For example, it is possible to have $v_x= 2e^{-2r_2}$ and $v_y=2$ where
the product $v_x v_y = 4 e^{-2r_2}$. If the resource variance
$e^{-2r_2} < 1/4$ (greater than \SI{6}{\dB}), then $v_x v_y < 1$,
surpassing what is sometimes called the
standard quantum limit.

For $w_y < w_x$, the optimal use of the probe is to set $\phi_2=0$ and
the optimal measurement is similar to Fig.~\ref{fig:4} but with
the measurements $X$ and $Y$ swapped. Repeating as before, we get
\begin{align}
  \label{eq:1}
 v_x=\frac{e^{-2r_2}}{t}\;\text{ and }\;  v_y=\frac{1}{1-t}
\end{align}
which is optimal when  $\frac{1}{1+e^{r_2}} \leq t \leq 1$. In terms
of the precisions, we have the relation
\begin{align}
  \label{eq:sxy1}
  \frac{ e^{-2r_2}}{v_x}+  \frac{1}{v_y}  \leq 1
\end{align}
which is plotted as the dotted grey line in Fig.~\ref{fig:3} for
$e^{-2r_2}=1/4$.

Finally to access the remaining region when $w_x=w_y$, we require
$t=\frac{1}{1+e^{r_2}}$ and the squeezing angle $\phi_2$ to vary
between $0$ and $\pi/2$. The optimal measurement is similar to
Fig.~\ref{fig:4} except that the quadrature measurement angles are 
set to $\phi_2+\pi/2$ in the upper arm  and $\phi_2$ in the lower
arm. Each of the measurement carry information on both $\theta_x$
and $\theta_y$. The two measurement outcomes, denoted by random
variables $M_1$ and $M_2$, follow Gaussian distributions with
\begin{align}
  \label{eq:2}
  \text{mean}(M_1) &= \sqrt{1-t}\left( \theta_y \cos \phi_2 - \theta_x
                     \sin \phi_2 \right)\;,\\
  \text{var}(M_1) &=1\;,
\end{align}
and
\begin{align}
  \label{eq:3}
  \text{mean}(M_2) &=\sqrt{t}\left( \theta_y \sin\phi_2 + \theta_x \cos\phi_2\right)\;,\\
  \text{var}(M_2) &=e^{-2r_2}\;.
\end{align}
With this, we can form two unbiased estimators for $\theta_x$ and
$\theta_y$:
\begin{align}
  \label{eq:3}
  \hat{\theta}_x &=\frac{M_2 \cos\phi_2}{\sqrt{t}}-\frac{M_1 \sin \phi_2}{\sqrt{1-t}}\;,\\
  \hat{\theta}_y &=\frac{M_2 \sin\phi_2}{\sqrt{t}}+\frac{M_1 \cos \phi_2}{\sqrt{1-t}}\;.
\end{align}
The variances of these estimators are
\begin{align}
  \label{eq:4}
  \text{var}(\hat{\theta}_x) &= \frac{e^{-2r_2}\cos^2\phi_2}{t}+
                               \frac{\sin^2 \phi_2}{1-t}\\
  &=\left( 1+e^{r_2}\right)e^{-2r_2}\left(\cos^2\phi_2+e^{r_2}\sin^2\phi_2 \right)\;,
\end{align}
and
\begin{align}
  \text{var}(\hat{\theta}_y) &= \frac{e^{-2r_2}\sin^2\phi_2}{t}+
                               \frac{\cos^2 \phi_2}{1-t}\\
  &=\left( 1+e^{r_2}\right)e^{-2r_2}\left(\sin^2\phi_2+e^{r_2}\cos^2\phi_2 \right)\;,
\end{align}
which saturates the bound~(\ref{eq:wxewy}).
\begin{figure}[t]
    \centering
\includegraphics[width=0.7\columnwidth]{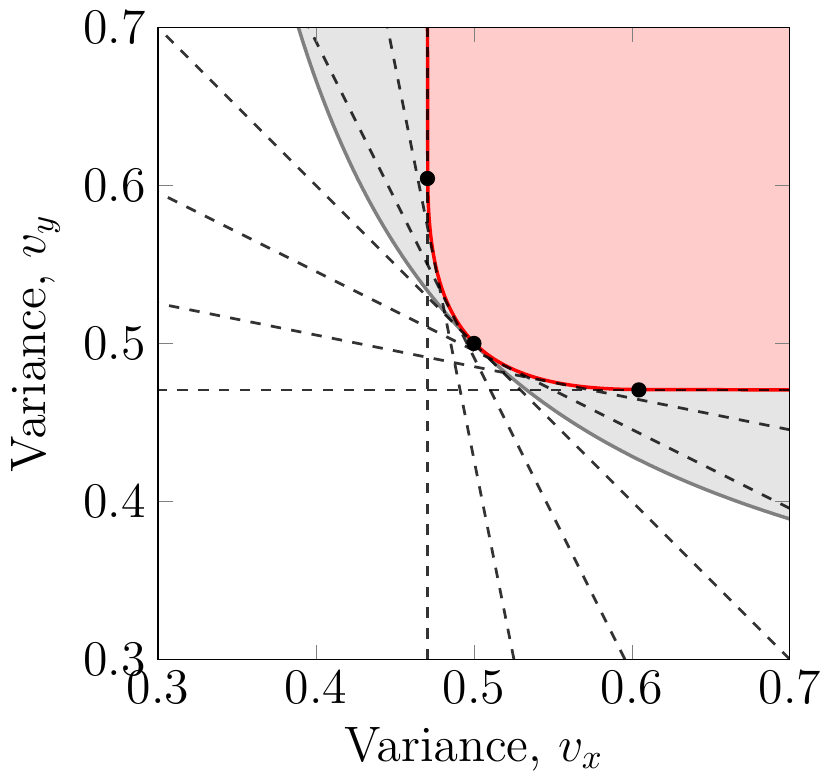}
\caption{
  \label{f:twoA_reg}Precision limits with two \SI{6}{dB} squeezed
  resource. Each black dashed line is a Holevo-CRB~(\ref{eq:5})
  determined by a value of $w_x$ and $w_y$ for a specific probe where $t=0.5$. The
  Holevo-CRB is an attainable bound, which means that for each of
  this line, there is a measurement that can reach at least one point
  on it. The three
  dots corresponds to the three special cases discussed in the main
  text in Eq.~(\ref{eq:3specialCase}). The red line, which is the collection of all the black
  line bounds, gives the achievable variances for this probe. The grey shaded area, defined
  by Eq.~(\ref{twoB}) is the collection of all accessible regions we
  can attain by varying $t$. We see that the red region touches the
  grey line at only one point when $v_x=v_y$. To reach the other
  points on the grey line, we need to use the resource in a different
  way with $t\neq0.5$.}
\end{figure}

\subsection{Example 2: Two equally squeezed state}
\begin{figure}[t]
    \centering
\includegraphics[width=0.8\columnwidth]{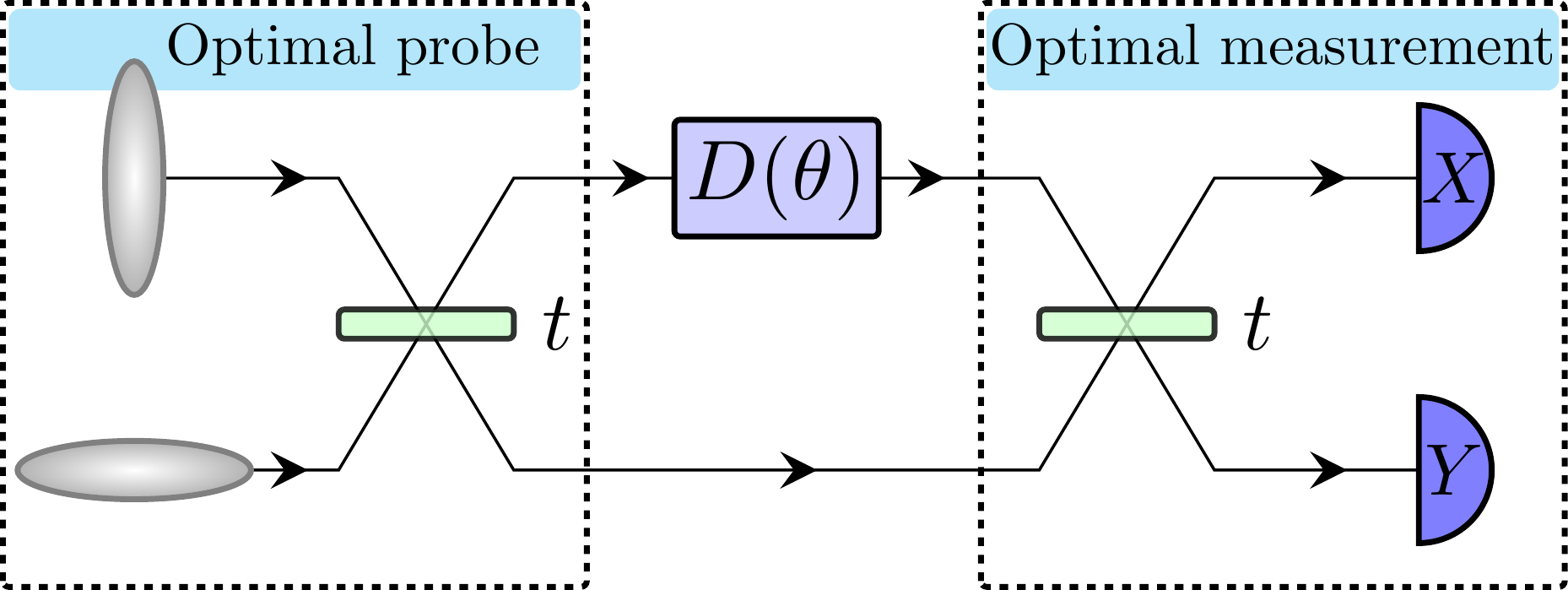}
\caption{
  \label{f:schBd} When $r_1=r_2$, for a fixed $w_x$ and $w_y$, the
  optimal probe that
    saturates the Holevo-CR
 bound is obtained by mixing the two squeezed states on a beam-splitter with  $t$ set to
 $\frac{\sqrt{w_y}}{\sqrt{w_x}+\sqrt{w_y}}$.
 The optimal measurement is to
 disentangle the probe into a product of single-mode
 states and measure $X$ on
  the first mode and $Y$ on the second mode. This gives the
  variances in Eq.~(\ref{bd_var}).}
\end{figure}
In our second example, we walk through the derivations of our main
result in the special case where the initial resource are two squeezed
states having an equal amount of squeezing $r_1=r_2=r$. In this case,
when $\phi_2=\phi_1+\pi/2$, the Holevo-CRB can be simplified to
\begin{align}
  \label{eq:5}
  w_x v_x+w_y v_y \geq f_\text{HCR}=\min_\lambda \{w_x f_x + w_y f_y \}\;,
\end{align}
where
\begin{align}
  f_x&\coloneqq \frac{\left(1+\lambda\sqrt{t}e^r \right)^2+\lambda^2 (1-t)e^{-2r}}{\left(\lambda+\sqrt{t}e^{-r}\right)^2}\;,\\
  f_y&\coloneqq \frac{\left(1+\lambda\sqrt{t}e^r \right)^2+\lambda^2 (1-t)e^{-2r}}{(1-t)e^{-2r}}\;.
\end{align}
In general, there is no analytical solution for the optimal value of
$\lambda$. To see how this leads to the main result in
Eq.~(\ref{eq:3a}), let us first consider a specific use of the
resource by interfering the two squeezed states on a beam splitter
with $t=0.5$ as shown in Fig.~\ref{fig:2a}. In this case, the
optimal $ \lambda $ that minimises $f_\text{HCR}$ is given by
$ \lambda^* ={-} e^{-r} (1+\gamma)/\sqrt{2}$ where $\gamma$ is the
positive solution to the quartic equation
\begin{align}
  \frac{w_y}{w_x} \gamma^3( \gamma-\tanh 2r)+\gamma \tanh 2r -1=0\;.
\end{align}
We can solve some special cases analytically:
\begin{alignat}{2}
  \label{eq:3specialCase}
( w_x=w_y=1)&:\;  v_x + v_y \geq 4 e^{-2r} &\text{ at }
\lambda^*&=-\sqrt{2}e^{-r}\nonumber \\
(w_x=1, w_y=0)&:\; v_x \geq  \frac{1}{\cosh 2r} &\text{ at }   \lambda^*&= \frac{-e^r}{\sqrt{2}\sinh 2r} \nonumber\\
(w_x=0, w_y=1)&:\;  v_y \geq  \frac{1}{\cosh 2r} &\text{ at }   \lambda^*&=\frac{-e^r}{\sqrt{2}\cosh 2r}\,.
\end{alignat}
For other values of $w_x/w_y$, $\lambda^*$ can be calculated
numerically and several of these bounds are plotted as the dashed lines in
Fig.~\ref{f:twoA_reg} when $e^{-2r}=1/4$. The envelope of these bounds is defined by the
parametric equation $v_x=f_x$ and $v_y=f_y$ for
${-}\frac{e^r}{\sqrt{2}\sinh 2r} < \lambda <
{-}\frac{e^r}{\sqrt{2}\cosh{2r}}$ and by construction can always be
reached. This is the precision limit attainable by the probe and is
plotted in red in Fig.~\ref{f:twoA_reg}. It is interesting to note
that the optimal variance of $v_x=\frac{1}{\cosh 2r}$ can be
achieved for any $v_y \geq \frac{\cosh 2r}{\sinh^2 2r}$.

The optimal precision as given by Eq.~(\ref{eq:3a}) is plotted in grey
in Fig.~\ref{f:twoA_reg}. We see that setting $t=0.5$ is only optimal
when $w_x=w_y$ which gives
$v_x=v_y=2e^{-2r}$~\cite{Steinlechner2013}. For every other points on
the grey line, a different probe configuration is needed to achieve
it. In other words, assigning different weights to the precisions of
the two quadratures will require the resource to be used
differently. In the extreme case where we are interested in only one
quadrature, the optimal scheme would be to just use one mode to sense
the displacement, as in squeezed state
interferometry~\cite{Caves1981,Xiao1987,Grangier1987}. In general,
when $w_x \neq w_y$, the optimal way to use the available resource is
to mix the two squeezed states on an unbalanced beam-splitter with
transmissivity $t^*=\frac{\sqrt{w_y}}{\sqrt{w_x}+\sqrt{w_y}}$.  At
this value of $t$, $f_\text{HCR}$ in Eq.~(\ref{eq:5}) is minimised
when $ \lambda^* ={-}{e^{-r}}/{\sqrt{t^*}}$ which gives Holevo-CRB as
\begin{align}
 f_\text{HCR}
  =\left( \sqrt{w_x}+\sqrt{w_y} \right)^2 e^{-2r}\;.
\end{align}

The measurement that saturates this bound is shown in
Fig.~\ref{f:schBd}. After the second beam-splitter, the displaced
two-mode probe is separated into two independent single-mode probes
with displacements $\sqrt{1-t^*}\theta$ and
$\sqrt{t^*}\theta$. Measuring $X$ on the first mode and $Y$ on the
second gives
\begin{align}
  \label{bd_var}
  v_x= \frac{e^{-2r}}{1-t^*}\;\text{ and }\;v_y=\frac{e^{-2r}}{t^*}\;.
\end{align}
Upon eliminating $t^*$, we have
\begin{align}
  \label{twoB}
  \frac{1}{v_x}+ \frac{1}{v_y} = e^{2r}\;,
\end{align}
which saturates the bound~(\ref{eq:3a}). This precision relation
  quantifies the resource apportioning principle and implies that
the quantum resource available through the squeezed states has to be shared
between the two conjugate quadratures~\cite{Liu2019}. The effects of
channel noise and inefficient detectors are presented in the
supplementary materials.

\section{Discussions and conclusion}
\label{sec:discussions}
To summarise, we find precision bounds in simultaneous estimation of
two conjugate quadratures. These bounds quantify a resource
  apportioning principle that limits how much precision is achievable
  with a given resource. While we restrict to pure states and
two-mode states in this work to derive transparent analytical results,
our formalism can be generalised to mixed and multi-mode Gaussian
probes. These results can be applied to channel estimation when the
amplitude and phase displacements have different strengths. For
example, the phase signal can be much weaker than the amplitude signal
we are trying to detect. This problem can also be formulated in a
resource theory
framework~\cite{Idel2016,Takagi2018,Albarelli2018,Yadin2018,Zhuang2018,Kwon2018},
where squeezing is a resource and passive transformations are free
operations. In this framework, the monotone that quantifies the
value of the resource will depend on the weights $w_x$ and $w_y$
assigned to each parameter. What {\em optimal} means must depend on
the application which assigns the weights $w_x$ and $w_y$.

\section*{Acknowledgements}
We thank H. Jeng for help in deriving the proofs. This work was
supported in part by National Natural Science Foundation of China
(91836302, 91736105, 11527808) and National Key Research and
Development Program of China (2016YFA0301403). S.A. and P.K.L. is
supported by the Australian Research Council (ARC) under the Centre of
Excellence for Quantum Computation and Communication Technology
(CE110001027).


\begin{thebibliography}{55}%
\makeatletter
\providecommand \@ifxundefined [1]{%
 \@ifx{#1\undefined}
}%
\providecommand \@ifnum [1]{%
 \ifnum #1\expandafter \@firstoftwo
 \else \expandafter \@secondoftwo
 \fi
}%
\providecommand \@ifx [1]{%
 \ifx #1\expandafter \@firstoftwo
 \else \expandafter \@secondoftwo
 \fi
}%
\providecommand \natexlab [1]{#1}%
\providecommand \enquote  [1]{``#1''}%
\providecommand \bibnamefont  [1]{#1}%
\providecommand \bibfnamefont [1]{#1}%
\providecommand \citenamefont [1]{#1}%
\providecommand \href@noop [0]{\@secondoftwo}%
\providecommand \href [0]{\begingroup \@sanitize@url \@href}%
\providecommand \@href[1]{\@@startlink{#1}\@@href}%
\providecommand \@@href[1]{\endgroup#1\@@endlink}%
\providecommand \@sanitize@url [0]{\catcode `\\12\catcode `\$12\catcode
  `\&12\catcode `\#12\catcode `\^12\catcode `\_12\catcode `\%12\relax}%
\providecommand \@@startlink[1]{}%
\providecommand \@@endlink[0]{}%
\providecommand \url  [0]{\begingroup\@sanitize@url \@url }%
\providecommand \@url [1]{\endgroup\@href {#1}{\urlprefix }}%
\providecommand \urlprefix  [0]{URL }%
\providecommand \Eprint [0]{\href }%
\providecommand \doibase [0]{http://dx.doi.org/}%
\providecommand \selectlanguage [0]{\@gobble}%
\providecommand \bibinfo  [0]{\@secondoftwo}%
\providecommand \bibfield  [0]{\@secondoftwo}%
\providecommand \translation [1]{[#1]}%
\providecommand \BibitemOpen [0]{}%
\providecommand \bibitemStop [0]{}%
\providecommand \bibitemNoStop [0]{.\EOS\space}%
\providecommand \EOS [0]{\spacefactor3000\relax}%
\providecommand \BibitemShut  [1]{\csname bibitem#1\endcsname}%
\let\auto@bib@innerbib\@empty
\bibitem [{\citenamefont {Caves}(1981)}]{Caves1981}%
  \BibitemOpen
  \bibfield  {author} {\bibinfo {author} {\bibfnamefont {Carlton~M.}\
  \bibnamefont {Caves}},\ }\bibfield  {title} {\enquote {\bibinfo {title}
  {Quantum-mechanical noise in an interferometer},}\ }\href {\doibase
  10.1103/PhysRevD.23.1693} {\bibfield  {journal} {\bibinfo  {journal} {Phys.
  Rev. D}\ }\textbf {\bibinfo {volume} {23}},\ \bibinfo {pages} {1693--1708}
  (\bibinfo {year} {1981})}\BibitemShut {NoStop}%
\bibitem [{\citenamefont {Xiao}\ \emph {et~al.}(1987)\citenamefont {Xiao},
  \citenamefont {Wu},\ and\ \citenamefont {Kimble}}]{Xiao1987}%
  \BibitemOpen
  \bibfield  {author} {\bibinfo {author} {\bibfnamefont {Min}\ \bibnamefont
  {Xiao}}, \bibinfo {author} {\bibfnamefont {Ling-An}\ \bibnamefont {Wu}}, \
  and\ \bibinfo {author} {\bibfnamefont {H.~J.}\ \bibnamefont {Kimble}},\
  }\bibfield  {title} {\enquote {\bibinfo {title} {Precision measurement beyond
  the shot-noise limit},}\ }\href {\doibase 10.1103/PhysRevLett.59.278}
  {\bibfield  {journal} {\bibinfo  {journal} {Phys. Rev. Lett.}\ }\textbf
  {\bibinfo {volume} {59}},\ \bibinfo {pages} {278--281} (\bibinfo {year}
  {1987})}\BibitemShut {NoStop}%
\bibitem [{\citenamefont {Grangier}\ \emph {et~al.}(1987)\citenamefont
  {Grangier}, \citenamefont {Slusher}, \citenamefont {Yurke},\ and\
  \citenamefont {LaPorta}}]{Grangier1987}%
  \BibitemOpen
  \bibfield  {author} {\bibinfo {author} {\bibfnamefont {P.}~\bibnamefont
  {Grangier}}, \bibinfo {author} {\bibfnamefont {R.~E.}\ \bibnamefont
  {Slusher}}, \bibinfo {author} {\bibfnamefont {B.}~\bibnamefont {Yurke}}, \
  and\ \bibinfo {author} {\bibfnamefont {A.}~\bibnamefont {LaPorta}},\
  }\bibfield  {title} {\enquote {\bibinfo {title} {Squeezed-light--enhanced
  polarization interferometer},}\ }\href {\doibase 10.1103/PhysRevLett.59.2153}
  {\bibfield  {journal} {\bibinfo  {journal} {Phys. Rev. Lett.}\ }\textbf
  {\bibinfo {volume} {59}},\ \bibinfo {pages} {2153--2156} (\bibinfo {year}
  {1987})}\BibitemShut {NoStop}%
\bibitem [{\citenamefont {D'Ariano}\ \emph {et~al.}(2001)\citenamefont
  {D'Ariano}, \citenamefont {Lo~Presti},\ and\ \citenamefont
  {Paris}}]{Dariano2001}%
  \BibitemOpen
  \bibfield  {author} {\bibinfo {author} {\bibfnamefont {G.~Mauro}\
  \bibnamefont {D'Ariano}}, \bibinfo {author} {\bibfnamefont {Paoloplacido}\
  \bibnamefont {Lo~Presti}}, \ and\ \bibinfo {author} {\bibfnamefont {Matteo
  G.~A.}\ \bibnamefont {Paris}},\ }\bibfield  {title} {\enquote {\bibinfo
  {title} {Using entanglement improves the precision of quantum
  measurements},}\ }\href {\doibase 10.1103/PhysRevLett.87.270404} {\bibfield
  {journal} {\bibinfo  {journal} {Phys. Rev. Lett.}\ }\textbf {\bibinfo
  {volume} {87}},\ \bibinfo {pages} {270404} (\bibinfo {year}
  {2001})}\BibitemShut {NoStop}%
\bibitem [{\citenamefont {Fujiwara}(2001)}]{Fujiwara2001}%
  \BibitemOpen
  \bibfield  {author} {\bibinfo {author} {\bibfnamefont {Akio}\ \bibnamefont
  {Fujiwara}},\ }\bibfield  {title} {\enquote {\bibinfo {title} {Quantum
  channel identification problem},}\ }\href {\doibase
  10.1103/PhysRevA.63.042304} {\bibfield  {journal} {\bibinfo  {journal} {Phys.
  Rev. A}\ }\textbf {\bibinfo {volume} {63}},\ \bibinfo {pages} {042304}
  (\bibinfo {year} {2001})}\BibitemShut {NoStop}%
\bibitem [{\citenamefont {Fischer}\ \emph {et~al.}(2001)\citenamefont
  {Fischer}, \citenamefont {Mack}, \citenamefont {Cirone},\ and\ \citenamefont
  {Freyberger}}]{Fischer2001}%
  \BibitemOpen
  \bibfield  {author} {\bibinfo {author} {\bibfnamefont {Dietmar~G.}\
  \bibnamefont {Fischer}}, \bibinfo {author} {\bibfnamefont {Holger}\
  \bibnamefont {Mack}}, \bibinfo {author} {\bibfnamefont {Markus~A.}\
  \bibnamefont {Cirone}}, \ and\ \bibinfo {author} {\bibfnamefont {Matthias}\
  \bibnamefont {Freyberger}},\ }\bibfield  {title} {\enquote {\bibinfo {title}
  {Enhanced estimation of a noisy quantum channel using entanglement},}\ }\href
  {\doibase 10.1103/PhysRevA.64.022309} {\bibfield  {journal} {\bibinfo
  {journal} {Phys. Rev. A}\ }\textbf {\bibinfo {volume} {64}},\ \bibinfo
  {pages} {022309} (\bibinfo {year} {2001})}\BibitemShut {NoStop}%
\bibitem [{\citenamefont {Sasaki}\ \emph {et~al.}(2002)\citenamefont {Sasaki},
  \citenamefont {Ban},\ and\ \citenamefont {Barnett}}]{Sasaki2002}%
  \BibitemOpen
  \bibfield  {author} {\bibinfo {author} {\bibfnamefont {Masahide}\
  \bibnamefont {Sasaki}}, \bibinfo {author} {\bibfnamefont {Masashi}\
  \bibnamefont {Ban}}, \ and\ \bibinfo {author} {\bibfnamefont {Stephen~M.}\
  \bibnamefont {Barnett}},\ }\bibfield  {title} {\enquote {\bibinfo {title}
  {Optimal parameter estimation of a depolarizing channel},}\ }\href {\doibase
  10.1103/PhysRevA.66.022308} {\bibfield  {journal} {\bibinfo  {journal} {Phys.
  Rev. A}\ }\textbf {\bibinfo {volume} {66}},\ \bibinfo {pages} {022308}
  (\bibinfo {year} {2002})}\BibitemShut {NoStop}%
\bibitem [{\citenamefont {Fujiwara}\ and\ \citenamefont
  {Imai}(2003)}]{Fujiwara2003}%
  \BibitemOpen
  \bibfield  {author} {\bibinfo {author} {\bibfnamefont {Akio}\ \bibnamefont
  {Fujiwara}}\ and\ \bibinfo {author} {\bibfnamefont {Hiroshi}\ \bibnamefont
  {Imai}},\ }\bibfield  {title} {\enquote {\bibinfo {title} {Quantum parameter
  estimation of a generalized pauli channel},}\ }\href {\doibase
  10.1088/0305-4470/36/29/314} {\bibfield  {journal} {\bibinfo  {journal}
  {Journal of Physics A: Mathematical and General}\ }\textbf {\bibinfo {volume}
  {36}},\ \bibinfo {pages} {8093–8103} (\bibinfo {year} {2003})}\BibitemShut
  {NoStop}%
\bibitem [{\citenamefont {Ballester}(2004)}]{Ballester2004}%
  \BibitemOpen
  \bibfield  {author} {\bibinfo {author} {\bibfnamefont {Manuel~A.}\
  \bibnamefont {Ballester}},\ }\bibfield  {title} {\enquote {\bibinfo {title}
  {Estimation of unitary quantum operations},}\ }\href {\doibase
  10.1103/PhysRevA.69.022303} {\bibfield  {journal} {\bibinfo  {journal} {Phys.
  Rev. A}\ }\textbf {\bibinfo {volume} {69}},\ \bibinfo {pages} {022303}
  (\bibinfo {year} {2004})}\BibitemShut {NoStop}%
\bibitem [{\citenamefont {Giovannetti}\ \emph {et~al.}(2004)\citenamefont
  {Giovannetti}, \citenamefont {Lloyd},\ and\ \citenamefont
  {Maccone}}]{Giovannetti2004}%
  \BibitemOpen
  \bibfield  {author} {\bibinfo {author} {\bibfnamefont {Vittorio}\
  \bibnamefont {Giovannetti}}, \bibinfo {author} {\bibfnamefont {Seth}\
  \bibnamefont {Lloyd}}, \ and\ \bibinfo {author} {\bibfnamefont {Lorenzo}\
  \bibnamefont {Maccone}},\ }\bibfield  {title} {\enquote {\bibinfo {title}
  {Quantum-enhanced measurements: beating the standard quantum limit},}\ }\href
  {\doibase 10.1126/science.1104149} {\bibfield  {journal} {\bibinfo  {journal}
  {Science}\ }\textbf {\bibinfo {volume} {306}},\ \bibinfo {pages} {1330}
  (\bibinfo {year} {2004})}\BibitemShut {NoStop}%
\bibitem [{\citenamefont {Genoni}\ \emph {et~al.}(2013)\citenamefont {Genoni},
  \citenamefont {Paris}, \citenamefont {Adesso}, \citenamefont {Nha},
  \citenamefont {Knight},\ and\ \citenamefont {Kim}}]{Genoni2013}%
  \BibitemOpen
  \bibfield  {author} {\bibinfo {author} {\bibfnamefont {MG}~\bibnamefont
  {Genoni}}, \bibinfo {author} {\bibfnamefont {MGA}\ \bibnamefont {Paris}},
  \bibinfo {author} {\bibfnamefont {G}~\bibnamefont {Adesso}}, \bibinfo
  {author} {\bibfnamefont {H}~\bibnamefont {Nha}}, \bibinfo {author}
  {\bibfnamefont {PL}~\bibnamefont {Knight}}, \ and\ \bibinfo {author}
  {\bibfnamefont {MS}~\bibnamefont {Kim}},\ }\bibfield  {title} {\enquote
  {\bibinfo {title} {Optimal estimation of joint parameters in phase space},}\
  }\href {http://pra.aps.org/abstract/PRA/v87/i1/e012107} {\bibfield  {journal}
  {\bibinfo  {journal} {Phys. Rev. A}\ }\textbf {\bibinfo {volume} {87}},\
  \bibinfo {pages} {012107} (\bibinfo {year} {2013})}\BibitemShut {NoStop}%
\bibitem [{\citenamefont {Rigovacca}\ \emph {et~al.}(2017)\citenamefont
  {Rigovacca}, \citenamefont {Farace}, \citenamefont {Souza}, \citenamefont
  {{De Pasquale}}, \citenamefont {Giovannetti},\ and\ \citenamefont
  {Adesso}}]{Rigovacca2017}%
  \BibitemOpen
  \bibfield  {author} {\bibinfo {author} {\bibfnamefont {Luca}\ \bibnamefont
  {Rigovacca}}, \bibinfo {author} {\bibfnamefont {Alessandro}\ \bibnamefont
  {Farace}}, \bibinfo {author} {\bibfnamefont {Leonardo A.~M.}\ \bibnamefont
  {Souza}}, \bibinfo {author} {\bibfnamefont {Antonella}\ \bibnamefont {{De
  Pasquale}}}, \bibinfo {author} {\bibfnamefont {Vittorio}\ \bibnamefont
  {Giovannetti}}, \ and\ \bibinfo {author} {\bibfnamefont {Gerardo}\
  \bibnamefont {Adesso}},\ }\bibfield  {title} {\enquote {\bibinfo {title}
  {Versatile gaussian probes for squeezing estimation},}\ }\href {\doibase
  10.1103/PhysRevA.95.052331} {\bibfield  {journal} {\bibinfo  {journal} {Phys.
  Rev. A}\ }\textbf {\bibinfo {volume} {95}},\ \bibinfo {pages} {052331}
  (\bibinfo {year} {2017})}\BibitemShut {NoStop}%
\bibitem [{\citenamefont {Bradshaw}\ \emph {et~al.}(2017)\citenamefont
  {Bradshaw}, \citenamefont {Assad},\ and\ \citenamefont {Lam}}]{Bradshaw2017}%
  \BibitemOpen
  \bibfield  {author} {\bibinfo {author} {\bibfnamefont {Mark}\ \bibnamefont
  {Bradshaw}}, \bibinfo {author} {\bibfnamefont {Syed~M.}\ \bibnamefont
  {Assad}}, \ and\ \bibinfo {author} {\bibfnamefont {Ping~Koy}\ \bibnamefont
  {Lam}},\ }\bibfield  {title} {\enquote {\bibinfo {title} {A tight
  {C}ramér–{R}ao bound for joint parameter estimation with a pure two-mode
  squeezed probe},}\ }\href {\doibase 10.1016/j.physleta.2017.06.024}
  {\bibfield  {journal} {\bibinfo  {journal} {Physics Letters A}\ }\textbf
  {\bibinfo {volume} {381}},\ \bibinfo {pages} {2598–2607} (\bibinfo {year}
  {2017})}\BibitemShut {NoStop}%
\bibitem [{\citenamefont {Bradshaw}\ \emph {et~al.}(2018)\citenamefont
  {Bradshaw}, \citenamefont {Lam},\ and\ \citenamefont {Assad}}]{Bradshaw2018}%
  \BibitemOpen
  \bibfield  {author} {\bibinfo {author} {\bibfnamefont {Mark}\ \bibnamefont
  {Bradshaw}}, \bibinfo {author} {\bibfnamefont {Ping~Koy}\ \bibnamefont
  {Lam}}, \ and\ \bibinfo {author} {\bibfnamefont {Syed~M.}\ \bibnamefont
  {Assad}},\ }\bibfield  {title} {\enquote {\bibinfo {title} {Ultimate
  precision of joint quadrature parameter estimation with a gaussian probe},}\
  }\href {\doibase 10.1103/physreva.97.012106} {\bibfield  {journal} {\bibinfo
  {journal} {Phys. Rev. A}\ }\textbf {\bibinfo {volume} {97}},\ \bibinfo
  {pages} {012106} (\bibinfo {year} {2018})}\BibitemShut {NoStop}%
\bibitem [{\citenamefont {Liu}\ \emph {et~al.}(2018)\citenamefont {Liu},
  \citenamefont {Li}, \citenamefont {Cui}, \citenamefont {Huo}, \citenamefont
  {Assad}, \citenamefont {Li},\ and\ \citenamefont {Ou}}]{Liu2018}%
  \BibitemOpen
  \bibfield  {author} {\bibinfo {author} {\bibfnamefont {Yuhong}\ \bibnamefont
  {Liu}}, \bibinfo {author} {\bibfnamefont {Jiamin}\ \bibnamefont {Li}},
  \bibinfo {author} {\bibfnamefont {Liang}\ \bibnamefont {Cui}}, \bibinfo
  {author} {\bibfnamefont {Nan}\ \bibnamefont {Huo}}, \bibinfo {author}
  {\bibfnamefont {Syed~M.}\ \bibnamefont {Assad}}, \bibinfo {author}
  {\bibfnamefont {Xiaoying}\ \bibnamefont {Li}}, \ and\ \bibinfo {author}
  {\bibfnamefont {Z.~Y.}\ \bibnamefont {Ou}},\ }\bibfield  {title} {\enquote
  {\bibinfo {title} {Loss-tolerant quantum dense metrology with {SU(1,1)}
  interferometer},}\ }\href {\doibase 10.1364/oe.26.027705} {\bibfield
  {journal} {\bibinfo  {journal} {Opt. Express}\ }\textbf {\bibinfo {volume}
  {26}},\ \bibinfo {pages} {27705} (\bibinfo {year} {2018})}\BibitemShut
  {NoStop}%
\bibitem [{\citenamefont {Li}\ \emph {et~al.}(2018)\citenamefont {Li},
  \citenamefont {Liu}, \citenamefont {Cui}, \citenamefont {Huo}, \citenamefont
  {Assad}, \citenamefont {Li},\ and\ \citenamefont {Ou}}]{Li2018}%
  \BibitemOpen
  \bibfield  {author} {\bibinfo {author} {\bibfnamefont {Jiamin}\ \bibnamefont
  {Li}}, \bibinfo {author} {\bibfnamefont {Yuhong}\ \bibnamefont {Liu}},
  \bibinfo {author} {\bibfnamefont {Liang}\ \bibnamefont {Cui}}, \bibinfo
  {author} {\bibfnamefont {Nan}\ \bibnamefont {Huo}}, \bibinfo {author}
  {\bibfnamefont {Syed~M.}\ \bibnamefont {Assad}}, \bibinfo {author}
  {\bibfnamefont {Xiaoying}\ \bibnamefont {Li}}, \ and\ \bibinfo {author}
  {\bibfnamefont {Z.~Y.}\ \bibnamefont {Ou}},\ }\bibfield  {title} {\enquote
  {\bibinfo {title} {Joint measurement of multiple noncommuting parameters},}\
  }\href {\doibase 10.1103/physreva.97.052127} {\bibfield  {journal} {\bibinfo
  {journal} {Phys. Rev. A}\ }\textbf {\bibinfo {volume} {97}},\ \bibinfo
  {pages} {052127} (\bibinfo {year} {2018})}\BibitemShut {NoStop}%
\bibitem [{\citenamefont {Gupta}\ \emph {et~al.}(2018)\citenamefont {Gupta},
  \citenamefont {Schmittberger}, \citenamefont {Anderson}, \citenamefont
  {Jones},\ and\ \citenamefont {Lett}}]{Gupta2018}%
  \BibitemOpen
  \bibfield  {author} {\bibinfo {author} {\bibfnamefont {Prasoon}\ \bibnamefont
  {Gupta}}, \bibinfo {author} {\bibfnamefont {Bonnie~L.}\ \bibnamefont
  {Schmittberger}}, \bibinfo {author} {\bibfnamefont {Brian~E.}\ \bibnamefont
  {Anderson}}, \bibinfo {author} {\bibfnamefont {Kevin~M.}\ \bibnamefont
  {Jones}}, \ and\ \bibinfo {author} {\bibfnamefont {Paul~D.}\ \bibnamefont
  {Lett}},\ }\bibfield  {title} {\enquote {\bibinfo {title} {Optimized phase
  sensing in a truncated su(1,1) interferometer},}\ }\href {\doibase
  10.1364/OE.26.000391} {\bibfield  {journal} {\bibinfo  {journal} {Opt.
  Express}\ }\textbf {\bibinfo {volume} {26}},\ \bibinfo {pages} {391}
  (\bibinfo {year} {2018})}\BibitemShut {NoStop}%
\bibitem [{\citenamefont {Aasi}\ \emph {et~al.}(2013)\citenamefont {Aasi},
  \citenamefont {Abadie}, \citenamefont {Abbott}, \citenamefont {Abbott},
  \citenamefont {Abbott}, \citenamefont {Abernathy}, \citenamefont {Adams},
  \citenamefont {Adams}, \citenamefont {Addesso}, \citenamefont {Adhikari},\
  and\ \citenamefont {et~al.}}]{aasi2013}%
  \BibitemOpen
  \bibfield  {author} {\bibinfo {author} {\bibfnamefont {J.}~\bibnamefont
  {Aasi}}, \bibinfo {author} {\bibfnamefont {J.}~\bibnamefont {Abadie}},
  \bibinfo {author} {\bibfnamefont {B.~P.}\ \bibnamefont {Abbott}}, \bibinfo
  {author} {\bibfnamefont {R.}~\bibnamefont {Abbott}}, \bibinfo {author}
  {\bibfnamefont {T.~D.}\ \bibnamefont {Abbott}}, \bibinfo {author}
  {\bibfnamefont {M.~R.}\ \bibnamefont {Abernathy}}, \bibinfo {author}
  {\bibfnamefont {C.}~\bibnamefont {Adams}}, \bibinfo {author} {\bibfnamefont
  {T.}~\bibnamefont {Adams}}, \bibinfo {author} {\bibfnamefont
  {P.}~\bibnamefont {Addesso}}, \bibinfo {author} {\bibfnamefont {R.~X.}\
  \bibnamefont {Adhikari}}, \ and\ \bibinfo {author} {\bibnamefont {et~al.}},\
  }\bibfield  {title} {\enquote {\bibinfo {title} {Enhanced sensitivity of the
  ligo gravitational wave detector by using squeezed states of light},}\ }\href
  {\doibase 10.1038/nphoton.2013.177} {\bibfield  {journal} {\bibinfo
  {journal} {Nature Photonics}\ }\textbf {\bibinfo {volume} {7}},\ \bibinfo
  {pages} {613–619} (\bibinfo {year} {2013})}\BibitemShut {NoStop}%
\bibitem [{\citenamefont {Grote}\ \emph {et~al.}(2013)\citenamefont {Grote},
  \citenamefont {Danzmann}, \citenamefont {Dooley}, \citenamefont {Schnabel},
  \citenamefont {Slutsky},\ and\ \citenamefont {Vahlbruch}}]{grote2013}%
  \BibitemOpen
  \bibfield  {author} {\bibinfo {author} {\bibfnamefont {H.}~\bibnamefont
  {Grote}}, \bibinfo {author} {\bibfnamefont {K.}~\bibnamefont {Danzmann}},
  \bibinfo {author} {\bibfnamefont {K.~L.}\ \bibnamefont {Dooley}}, \bibinfo
  {author} {\bibfnamefont {R.}~\bibnamefont {Schnabel}}, \bibinfo {author}
  {\bibfnamefont {J.}~\bibnamefont {Slutsky}}, \ and\ \bibinfo {author}
  {\bibfnamefont {H.}~\bibnamefont {Vahlbruch}},\ }\bibfield  {title} {\enquote
  {\bibinfo {title} {First long-term application of squeezed states of light in
  a gravitational-wave observatory},}\ }\href {\doibase
  10.1103/PhysRevLett.110.181101} {\bibfield  {journal} {\bibinfo  {journal}
  {Phys. Rev. Lett.}\ }\textbf {\bibinfo {volume} {110}},\ \bibinfo {pages}
  {181101} (\bibinfo {year} {2013})}\BibitemShut {NoStop}%
\bibitem [{\citenamefont {Arthurs}\ and\ \citenamefont
  {Kelly}(1965)}]{Arthurs1965}%
  \BibitemOpen
  \bibfield  {author} {\bibinfo {author} {\bibfnamefont {E.}~\bibnamefont
  {Arthurs}}\ and\ \bibinfo {author} {\bibfnamefont {J.~L.}\ \bibnamefont
  {Kelly}},\ }\bibfield  {title} {\enquote {\bibinfo {title} {On the
  simultaneous measurement of a pair of conjugate observables},}\ }\href
  {\doibase 10.1002/j.1538-7305.1965.tb01684.x} {\bibfield  {journal} {\bibinfo
   {journal} {Bell System Technical Journal}\ }\textbf {\bibinfo {volume}
  {44}},\ \bibinfo {pages} {725–729} (\bibinfo {year} {1965})}\BibitemShut
  {NoStop}%
\bibitem [{\citenamefont {Yuen}(1982)}]{Yuen1982}%
  \BibitemOpen
  \bibfield  {author} {\bibinfo {author} {\bibfnamefont {Horace~P.}\
  \bibnamefont {Yuen}},\ }\bibfield  {title} {\enquote {\bibinfo {title}
  {Generalized quantum measurements and approximate simultaneous measurements
  of noncommuting observables},}\ }\href {\doibase
  10.1016/0375-9601(82)90359-0} {\bibfield  {journal} {\bibinfo  {journal}
  {Physics Letters A}\ }\textbf {\bibinfo {volume} {91}},\ \bibinfo {pages}
  {101–104} (\bibinfo {year} {1982})}\BibitemShut {NoStop}%
\bibitem [{\citenamefont {Arthurs}\ and\ \citenamefont
  {Goodman}(1988)}]{Arthurs1988}%
  \BibitemOpen
  \bibfield  {author} {\bibinfo {author} {\bibfnamefont {E.}~\bibnamefont
  {Arthurs}}\ and\ \bibinfo {author} {\bibfnamefont {M.~S.}\ \bibnamefont
  {Goodman}},\ }\bibfield  {title} {\enquote {\bibinfo {title} {Quantum
  correlations: A generalized heisenberg uncertainty relation},}\ }\href
  {\doibase 10.1103/physrevlett.60.2447} {\bibfield  {journal} {\bibinfo
  {journal} {Phys. Rev. Lett.}\ }\textbf {\bibinfo {volume} {60}},\ \bibinfo
  {pages} {2447–2449} (\bibinfo {year} {1988})}\BibitemShut {NoStop}%
\bibitem [{\citenamefont {Duivenvoorden}\ \emph {et~al.}(2017)\citenamefont
  {Duivenvoorden}, \citenamefont {Terhal},\ and\ \citenamefont
  {Weigand}}]{Duivenvoorden2017}%
  \BibitemOpen
  \bibfield  {author} {\bibinfo {author} {\bibfnamefont {Kasper}\ \bibnamefont
  {Duivenvoorden}}, \bibinfo {author} {\bibfnamefont {Barbara~M.}\ \bibnamefont
  {Terhal}}, \ and\ \bibinfo {author} {\bibfnamefont {Daniel}\ \bibnamefont
  {Weigand}},\ }\bibfield  {title} {\enquote {\bibinfo {title} {Single-mode
  displacement sensor},}\ }\href {\doibase 10.1103/physreva.95.012305}
  {\bibfield  {journal} {\bibinfo  {journal} {Phys. Rev. A}\ }\textbf {\bibinfo
  {volume} {95}},\ \bibinfo {pages} {2469--9934} (\bibinfo {year}
  {2017})}\BibitemShut {NoStop}%
\bibitem [{\citenamefont {Braunstein}\ and\ \citenamefont
  {Kimble}(2000)}]{Braunstein2000}%
  \BibitemOpen
  \bibfield  {author} {\bibinfo {author} {\bibfnamefont {Samuel~L.}\
  \bibnamefont {Braunstein}}\ and\ \bibinfo {author} {\bibfnamefont {H.~J.}\
  \bibnamefont {Kimble}},\ }\bibfield  {title} {\enquote {\bibinfo {title}
  {Dense coding for continuous variables},}\ }\href {\doibase
  10.1103/physreva.61.042302} {\bibfield  {journal} {\bibinfo  {journal} {Phys.
  Rev. A}\ }\textbf {\bibinfo {volume} {61}},\ \bibinfo {pages} {042302}
  (\bibinfo {year} {2000})}\BibitemShut {NoStop}%
\bibitem [{\citenamefont {Zhang}\ and\ \citenamefont {Peng}(2000)}]{Zhang2000}%
  \BibitemOpen
  \bibfield  {author} {\bibinfo {author} {\bibfnamefont {Jing}\ \bibnamefont
  {Zhang}}\ and\ \bibinfo {author} {\bibfnamefont {Kunchi}\ \bibnamefont
  {Peng}},\ }\bibfield  {title} {\enquote {\bibinfo {title} {Quantum
  teleportation and dense coding by means of bright amplitude-squeezed light
  and direct measurement of a bell state},}\ }\href {\doibase
  10.1103/physreva.62.064302} {\bibfield  {journal} {\bibinfo  {journal} {Phys.
  Rev. A}\ }\textbf {\bibinfo {volume} {62}},\ \bibinfo {pages} {064302}
  (\bibinfo {year} {2000})}\BibitemShut {NoStop}%
\bibitem [{\citenamefont {Li}\ \emph {et~al.}(2002)\citenamefont {Li},
  \citenamefont {Pan}, \citenamefont {Jing}, \citenamefont {Zhang},
  \citenamefont {Xie},\ and\ \citenamefont {Peng}}]{Li2002}%
  \BibitemOpen
  \bibfield  {author} {\bibinfo {author} {\bibfnamefont {Xiaoying}\
  \bibnamefont {Li}}, \bibinfo {author} {\bibfnamefont {Qing}\ \bibnamefont
  {Pan}}, \bibinfo {author} {\bibfnamefont {Jietai}\ \bibnamefont {Jing}},
  \bibinfo {author} {\bibfnamefont {Jing}\ \bibnamefont {Zhang}}, \bibinfo
  {author} {\bibfnamefont {Changde}\ \bibnamefont {Xie}}, \ and\ \bibinfo
  {author} {\bibfnamefont {Kunchi}\ \bibnamefont {Peng}},\ }\bibfield  {title}
  {\enquote {\bibinfo {title} {Quantum dense coding exploiting a bright
  einstein-podolsky-rosen beam},}\ }\href {\doibase
  10.1103/PhysRevLett.88.047904} {\bibfield  {journal} {\bibinfo  {journal}
  {Phys. Rev. Lett.}\ }\textbf {\bibinfo {volume} {88}},\ \bibinfo {pages}
  {047904} (\bibinfo {year} {2002})}\BibitemShut {NoStop}%
\bibitem [{\citenamefont {Steinlechner}\ \emph {et~al.}(2013)\citenamefont
  {Steinlechner}, \citenamefont {Bauchrowitz}, \citenamefont {Meinders},
  \citenamefont {Munro}, \citenamefont {Danzmann},\ and\ \citenamefont
  {Schnabel}}]{Steinlechner2013}%
  \BibitemOpen
  \bibfield  {author} {\bibinfo {author} {\bibfnamefont {Sebastian}\
  \bibnamefont {Steinlechner}}, \bibinfo {author} {\bibfnamefont {Joran}\
  \bibnamefont {Bauchrowitz}}, \bibinfo {author} {\bibfnamefont {Melanie}\
  \bibnamefont {Meinders}}, \bibinfo {author} {\bibfnamefont {Helge}\
  \bibnamefont {Munro}, \bibfnamefont {W~Jller-Ebhardt}}, \bibinfo {author}
  {\bibfnamefont {Karsten}\ \bibnamefont {Danzmann}}, \ and\ \bibinfo {author}
  {\bibfnamefont {Roman}\ \bibnamefont {Schnabel}},\ }\bibfield  {title}
  {\enquote {\bibinfo {title} {Quantum-dense metrology},}\ }\href {\doibase
  10.1038/nphoton.2013.150} {\bibfield  {journal} {\bibinfo  {journal} {Nature
  Photonics}\ }\textbf {\bibinfo {volume} {7}},\ \bibinfo {pages} {626–630}
  (\bibinfo {year} {2013})}\BibitemShut {NoStop}%
\bibitem [{\citenamefont {Helstrom}(1967)}]{Helstrom1967}%
  \BibitemOpen
  \bibfield  {author} {\bibinfo {author} {\bibfnamefont {CW}~\bibnamefont
  {Helstrom}},\ }\bibfield  {title} {\enquote {\bibinfo {title} {Minimum
  mean-squared error of estimates in quantum statistics},}\ }\href@noop {}
  {\bibfield  {journal} {\bibinfo  {journal} {Phys. Lett. A}\ }\textbf
  {\bibinfo {volume} {25}},\ \bibinfo {pages} {101–102} (\bibinfo {year}
  {1967})}\BibitemShut {NoStop}%
\bibitem [{\citenamefont {Helstrom}(1969)}]{Helstrom1969}%
  \BibitemOpen
  \bibfield  {author} {\bibinfo {author} {\bibfnamefont {Carl~W}\ \bibnamefont
  {Helstrom}},\ }\bibfield  {title} {\enquote {\bibinfo {title} {Quantum
  detection and estimation theory},}\ }\href@noop {} {\bibfield  {journal}
  {\bibinfo  {journal} {Journal of Statistical Physics}\ }\textbf {\bibinfo
  {volume} {1}},\ \bibinfo {pages} {231–252} (\bibinfo {year}
  {1969})}\BibitemShut {NoStop}%
\bibitem [{\citenamefont {Holevo}(1976)}]{Holevo1976}%
  \BibitemOpen
  \bibfield  {author} {\bibinfo {author} {\bibfnamefont {AS}~\bibnamefont
  {Holevo}},\ }\bibfield  {title} {\enquote {\bibinfo {title} {Noncommutative
  analogues of the {C}ramér-{R}ao inequality in the quantum measurement
  theory},}\ }in\ \href@noop {} {\emph {\bibinfo {booktitle} {Proceedings of
  the Third Japan—USSR Symposium on Probability Theory}}}\ (\bibinfo
  {organization} {Springer},\ \bibinfo {year} {1976})\ p.\ \bibinfo {pages}
  {194–222}\BibitemShut {NoStop}%
\bibitem [{\citenamefont {Holevo}(2011)}]{Holevo2011}%
  \BibitemOpen
  \bibfield  {author} {\bibinfo {author} {\bibfnamefont {Alexander~S}\
  \bibnamefont {Holevo}},\ }\href@noop {} {\emph {\bibinfo {title}
  {Probabilistic and statistical aspects of quantum theory}}},\ Vol.~\bibinfo
  {volume} {1}\ (\bibinfo  {publisher} {Springer Science \& Business Media},\
  \bibinfo {year} {2011})\BibitemShut {NoStop}%
\bibitem [{\citenamefont {Gao}\ and\ \citenamefont {Lee}(2014)}]{Gao2014}%
  \BibitemOpen
  \bibfield  {author} {\bibinfo {author} {\bibfnamefont {Yang}\ \bibnamefont
  {Gao}}\ and\ \bibinfo {author} {\bibfnamefont {Hwang}\ \bibnamefont {Lee}},\
  }\bibfield  {title} {\enquote {\bibinfo {title} {Bounds on quantum
  multiple-parameter estimation with gaussian state},}\ }\href {\doibase
  10.1140/epjd/e2014-50560-1} {\bibfield  {journal} {\bibinfo  {journal} {The
  European Physical Journal D}\ }\textbf {\bibinfo {volume} {68}},\ \bibinfo
  {pages} {1–7} (\bibinfo {year} {2014})}\BibitemShut {NoStop}%
\bibitem [{\citenamefont {Braunstein}\ and\ \citenamefont
  {Caves}(1994)}]{Braunstein1994}%
  \BibitemOpen
  \bibfield  {author} {\bibinfo {author} {\bibfnamefont {Samuel~L.}\
  \bibnamefont {Braunstein}}\ and\ \bibinfo {author} {\bibfnamefont
  {Carlton~M.}\ \bibnamefont {Caves}},\ }\bibfield  {title} {\enquote {\bibinfo
  {title} {Statistical distance and the geometry of quantum states},}\ }\href
  {\doibase 10.1103/physrevlett.72.3439} {\bibfield  {journal} {\bibinfo
  {journal} {Phys. Rev. Lett.}\ }\textbf {\bibinfo {volume} {72}},\ \bibinfo
  {pages} {3439–3443} (\bibinfo {year} {1994})}\BibitemShut {NoStop}%
\bibitem [{\citenamefont {Fujiwara}\ and\ \citenamefont
  {Nagaoka}(1995)}]{Fujiwara1995}%
  \BibitemOpen
  \bibfield  {author} {\bibinfo {author} {\bibfnamefont {Akio}\ \bibnamefont
  {Fujiwara}}\ and\ \bibinfo {author} {\bibfnamefont {Hiroshi}\ \bibnamefont
  {Nagaoka}},\ }\bibfield  {title} {\enquote {\bibinfo {title} {Quantum fisher
  metric and estimation for pure state models},}\ }\href {\doibase
  10.1016/0375-9601(95)00269-9} {\bibfield  {journal} {\bibinfo  {journal}
  {Phys. Lett. A}\ }\textbf {\bibinfo {volume} {201}},\ \bibinfo {pages}
  {119–124} (\bibinfo {year} {1995})}\BibitemShut {NoStop}%
\bibitem [{\citenamefont {Yuen}\ and\ \citenamefont {Lax}(1973)}]{Yuen1973}%
  \BibitemOpen
  \bibfield  {author} {\bibinfo {author} {\bibfnamefont {H}~\bibnamefont
  {Yuen}}\ and\ \bibinfo {author} {\bibfnamefont {Melvin}\ \bibnamefont
  {Lax}},\ }\bibfield  {title} {\enquote {\bibinfo {title} {Multiple-parameter
  quantum estimation and measurement of nonselfadjoint observables},}\
  }\href@noop {} {\bibfield  {journal} {\bibinfo  {journal} {IEEE Trans.
  Inform. Theory}\ }\textbf {\bibinfo {volume} {19}},\ \bibinfo {pages}
  {740–750} (\bibinfo {year} {1973})}\BibitemShut {NoStop}%
\bibitem [{\citenamefont {Belavkin}(1976)}]{Belavkin1976}%
  \BibitemOpen
  \bibfield  {author} {\bibinfo {author} {\bibfnamefont {Vyacheslav~P}\
  \bibnamefont {Belavkin}},\ }\bibfield  {title} {\enquote {\bibinfo {title}
  {Generalized uncertainty relations and efficient measurements in quantum
  systems},}\ }\href@noop {} {\bibfield  {journal} {\bibinfo  {journal}
  {Theoretical and Mathematical Physics}\ }\textbf {\bibinfo {volume} {26}},\
  \bibinfo {pages} {213–222} (\bibinfo {year} {1976})}\BibitemShut {NoStop}%
\bibitem [{\citenamefont {Fujiwara}(1994{\natexlab{a}})}]{Fujiwara1994}%
  \BibitemOpen
  \bibfield  {author} {\bibinfo {author} {\bibfnamefont {Akio}\ \bibnamefont
  {Fujiwara}},\ }\bibfield  {title} {\enquote {\bibinfo {title}
  {Multi-parameter pure state estimation based on the right logarithmic
  derivative},}\ }\href@noop {} {\bibfield  {journal} {\bibinfo  {journal}
  {Math. Eng. Tech. Rep}\ }\textbf {\bibinfo {volume} {94}},\ \bibinfo {pages}
  {94–10} (\bibinfo {year} {1994}{\natexlab{a}})}\BibitemShut {NoStop}%
\bibitem [{\citenamefont {Fujiwara}(1994{\natexlab{b}})}]{Fujiwara1994a}%
  \BibitemOpen
  \bibfield  {author} {\bibinfo {author} {\bibfnamefont {Akio}\ \bibnamefont
  {Fujiwara}},\ }\bibfield  {title} {\enquote {\bibinfo {title} {Linear random
  measurements of two non-commuting observables},}\ }\href@noop {} {\bibfield
  {journal} {\bibinfo  {journal} {Math. Eng. Tech. Rep}\ }\textbf {\bibinfo
  {volume} {94}} (\bibinfo {year} {1994}{\natexlab{b}})}\BibitemShut {NoStop}%
\bibitem [{\citenamefont {Fujiwara}\ and\ \citenamefont
  {Nagaoka}(1999)}]{Fujiwara1999}%
  \BibitemOpen
  \bibfield  {author} {\bibinfo {author} {\bibfnamefont {Akio}\ \bibnamefont
  {Fujiwara}}\ and\ \bibinfo {author} {\bibfnamefont {Hiroshi}\ \bibnamefont
  {Nagaoka}},\ }\bibfield  {title} {\enquote {\bibinfo {title} {An estimation
  theoretical characterization of coherent states},}\ }\href@noop {} {\bibfield
   {journal} {\bibinfo  {journal} {J. Math. Phys.}\ }\textbf {\bibinfo {volume}
  {40}},\ \bibinfo {pages} {4227–4239} (\bibinfo {year} {1999})}\BibitemShut
  {NoStop}%
\bibitem [{\citenamefont {Paris}(2009)}]{Paris2009}%
  \BibitemOpen
  \bibfield  {author} {\bibinfo {author} {\bibfnamefont {Matteo~GA}\
  \bibnamefont {Paris}},\ }\bibfield  {title} {\enquote {\bibinfo {title}
  {Quantum estimation for quantum technology},}\ }\href@noop {} {\bibfield
  {journal} {\bibinfo  {journal} {Int. J. Quantum Inf.}\ }\textbf {\bibinfo
  {volume} {7}},\ \bibinfo {pages} {125–137} (\bibinfo {year}
  {2009})}\BibitemShut {NoStop}%
\bibitem [{\citenamefont {Petz}\ and\ \citenamefont {Ghinea}(2011)}]{Petz2011}%
  \BibitemOpen
  \bibfield  {author} {\bibinfo {author} {\bibfnamefont {D.}~\bibnamefont
  {Petz}}\ and\ \bibinfo {author} {\bibfnamefont {C.}~\bibnamefont {Ghinea}},\
  }\enquote {\bibinfo {title} {Introduction to quantum fisher information},}\
  in\ \href {\doibase 10.1142/9789814338745_0015} {\emph {\bibinfo {booktitle}
  {Quantum Probability and Related Topics}}}\ (\bibinfo  {publisher} {World
  Scientific},\ \bibinfo {year} {2011})\ Chap.~\bibinfo {chapter} {15}, p.\
  \bibinfo {pages} {261–281}\BibitemShut {NoStop}%
\bibitem [{\citenamefont {Barndorff-Nielsen}\ and\ \citenamefont
  {Gill}(2000)}]{Barndorff-Nielsen2000}%
  \BibitemOpen
  \bibfield  {author} {\bibinfo {author} {\bibfnamefont {O~E}\ \bibnamefont
  {Barndorff-Nielsen}}\ and\ \bibinfo {author} {\bibfnamefont {R~D}\
  \bibnamefont {Gill}},\ }\bibfield  {title} {\enquote {\bibinfo {title}
  {Fisher information in quantum statistics},}\ }\href {\doibase
  10.1088/0305-4470/33/24/306} {\bibfield  {journal} {\bibinfo  {journal} {J.
  Phys. A: Math. Gen.}\ }\textbf {\bibinfo {volume} {33}},\ \bibinfo {pages}
  {4481–4490} (\bibinfo {year} {2000})}\BibitemShut {NoStop}%
\bibitem [{\citenamefont {Szczykulska}\ \emph {et~al.}(2016)\citenamefont
  {Szczykulska}, \citenamefont {Baumgratz},\ and\ \citenamefont
  {Datta}}]{Szczykulska2016}%
  \BibitemOpen
  \bibfield  {author} {\bibinfo {author} {\bibfnamefont {Magdalena}\
  \bibnamefont {Szczykulska}}, \bibinfo {author} {\bibfnamefont {Tillmann}\
  \bibnamefont {Baumgratz}}, \ and\ \bibinfo {author} {\bibfnamefont {Animesh}\
  \bibnamefont {Datta}},\ }\bibfield  {title} {\enquote {\bibinfo {title}
  {Multi-parameter quantum metrology},}\ }\href {\doibase
  10.1080/23746149.2016.1230476} {\bibfield  {journal} {\bibinfo  {journal}
  {Advances in Physics: X}\ }\textbf {\bibinfo {volume} {1}},\ \bibinfo {pages}
  {621–639} (\bibinfo {year} {2016})}\BibitemShut {NoStop}%
\bibitem [{\citenamefont {Suzuki}(2019)}]{Suzuki2019}%
  \BibitemOpen
  \bibfield  {author} {\bibinfo {author} {\bibfnamefont {Jun}\ \bibnamefont
  {Suzuki}},\ }\bibfield  {title} {\enquote {\bibinfo {title} {Information
  geometrical characterization of quantum statistical models in quantum
  estimation theory},}\ }\href {\doibase 10.3390/e21070703} {\bibfield
  {journal} {\bibinfo  {journal} {Entropy}\ }\textbf {\bibinfo {volume} {21}},\
  \bibinfo {pages} {703} (\bibinfo {year} {2019})}\BibitemShut {NoStop}%
\bibitem [{\citenamefont {Nagaoka}(2005)}]{Nagaoka2005}%
  \BibitemOpen
  \bibfield  {author} {\bibinfo {author} {\bibfnamefont {Hiroshi}\ \bibnamefont
  {Nagaoka}},\ }\bibfield  {title} {\enquote {\bibinfo {title} {A new approach
  to {C}ramér-{Rao} bounds for quantum state estimation},}\ }in\ \href
  {\doibase 10.1142/9789812563071_0009} {\emph {\bibinfo {booktitle}
  {Asymptotic Theory of Quantum Statistical Inference}}},\ \bibinfo {editor}
  {edited by\ \bibinfo {editor} {\bibfnamefont {Masahito}\ \bibnamefont
  {Hayashi}}}\ (\bibinfo  {publisher} {WORLD SCIENTIFIC},\ \bibinfo {year}
  {2005})\ p.\ \bibinfo {pages} {100–112}\BibitemShut {NoStop}%
\bibitem [{\citenamefont {Hayashi}(2006)}]{Hayashi2006}%
  \BibitemOpen
  \bibfield  {author} {\bibinfo {author} {\bibfnamefont {Masahito}\
  \bibnamefont {Hayashi}},\ }\href {\doibase 10.1007/3-540-30266-2} {\emph
  {\bibinfo {title} {Quantum Information An Introduction}}}\ (\bibinfo
  {publisher} {Springer Berlin Heidelberg},\ \bibinfo {year}
  {2006})\BibitemShut {NoStop}%
\bibitem [{\citenamefont {Yamagata}\ \emph {et~al.}(2013)\citenamefont
  {Yamagata}, \citenamefont {Fujiwara},\ and\ \citenamefont
  {Gill}}]{Yamagata2013}%
  \BibitemOpen
  \bibfield  {author} {\bibinfo {author} {\bibfnamefont {Koichi}\ \bibnamefont
  {Yamagata}}, \bibinfo {author} {\bibfnamefont {Akio}\ \bibnamefont
  {Fujiwara}}, \ and\ \bibinfo {author} {\bibfnamefont {Richard~D.}\
  \bibnamefont {Gill}},\ }\bibfield  {title} {\enquote {\bibinfo {title}
  {Quantum local asymptotic normality based on a new quantum likelihood
  ratio},}\ }\href {\doibase 10.1214/13-aos1147} {\bibfield  {journal}
  {\bibinfo  {journal} {The Annals of Statistics}\ }\textbf {\bibinfo {volume}
  {41}},\ \bibinfo {pages} {2197–2217} (\bibinfo {year} {2013})}\BibitemShut
  {NoStop}%
\bibitem [{\citenamefont {Liu}\ \emph {et~al.}(2019)\citenamefont {Liu},
  \citenamefont {Huo}, \citenamefont {Li}, \citenamefont {Cui}, \citenamefont
  {Li},\ and\ \citenamefont {Ou}}]{Liu2019}%
  \BibitemOpen
  \bibfield  {author} {\bibinfo {author} {\bibfnamefont {Yuhong}\ \bibnamefont
  {Liu}}, \bibinfo {author} {\bibfnamefont {Nan}\ \bibnamefont {Huo}}, \bibinfo
  {author} {\bibfnamefont {Jiamin}\ \bibnamefont {Li}}, \bibinfo {author}
  {\bibfnamefont {Liang}\ \bibnamefont {Cui}}, \bibinfo {author} {\bibfnamefont
  {Xiaoying}\ \bibnamefont {Li}}, \ and\ \bibinfo {author} {\bibfnamefont
  {Zheyu~Jeff}\ \bibnamefont {Ou}},\ }\bibfield  {title} {\enquote {\bibinfo
  {title} {Optimum quantum resource distribution for phase measurement and
  quantum information tapping in a dual-beam {SU(1,1)} interferometer},}\
  }\href {\doibase 10.1364/oe.27.011292} {\bibfield  {journal} {\bibinfo
  {journal} {Opt. Express}\ }\textbf {\bibinfo {volume} {27}},\ \bibinfo
  {pages} {11292} (\bibinfo {year} {2019})}\BibitemShut {NoStop}%
\bibitem [{\citenamefont {Olivares}\ and\ \citenamefont
  {Paris}(2011)}]{Olivares2011}%
  \BibitemOpen
  \bibfield  {author} {\bibinfo {author} {\bibfnamefont {Stefano}\ \bibnamefont
  {Olivares}}\ and\ \bibinfo {author} {\bibfnamefont {Matteo G.~A.}\
  \bibnamefont {Paris}},\ }\bibfield  {title} {\enquote {\bibinfo {title}
  {Fidelity matters: The birth of entanglement in the mixing of gaussian
  states},}\ }\href {\doibase 10.1103/physrevlett.107.170505} {\bibfield
  {journal} {\bibinfo  {journal} {Phys. Rev. Lett.}\ }\textbf {\bibinfo
  {volume} {107}},\ \bibinfo {pages} {170505} (\bibinfo {year}
  {2011})}\BibitemShut {NoStop}%
\bibitem [{\citenamefont {Idel}\ \emph {et~al.}(2016)\citenamefont {Idel},
  \citenamefont {Lercher},\ and\ \citenamefont {Wolf}}]{Idel2016}%
  \BibitemOpen
  \bibfield  {author} {\bibinfo {author} {\bibfnamefont {Martin}\ \bibnamefont
  {Idel}}, \bibinfo {author} {\bibfnamefont {Daniel}\ \bibnamefont {Lercher}},
  \ and\ \bibinfo {author} {\bibfnamefont {Michael~M}\ \bibnamefont {Wolf}},\
  }\bibfield  {title} {\enquote {\bibinfo {title} {An operational measure for
  squeezing},}\ }\href {\doibase 10.1088/1751-8113/49/44/445304} {\bibfield
  {journal} {\bibinfo  {journal} {Journal of Physics A: Mathematical and
  Theoretical}\ }\textbf {\bibinfo {volume} {49}},\ \bibinfo {pages} {445304}
  (\bibinfo {year} {2016})}\BibitemShut {NoStop}%
\bibitem [{\citenamefont {Takagi}\ and\ \citenamefont
  {Zhuang}(2018)}]{Takagi2018}%
  \BibitemOpen
  \bibfield  {author} {\bibinfo {author} {\bibfnamefont {Ryuji}\ \bibnamefont
  {Takagi}}\ and\ \bibinfo {author} {\bibfnamefont {Quntao}\ \bibnamefont
  {Zhuang}},\ }\bibfield  {title} {\enquote {\bibinfo {title} {Convex resource
  theory of non-gaussianity},}\ }\href {\doibase 10.1103/physreva.97.062337}
  {\bibfield  {journal} {\bibinfo  {journal} {Phys. Rev. A}\ }\textbf {\bibinfo
  {volume} {97}},\ \bibinfo {pages} {062337} (\bibinfo {year}
  {2018})}\BibitemShut {NoStop}%
\bibitem [{\citenamefont {Albarelli}\ \emph {et~al.}(2018)\citenamefont
  {Albarelli}, \citenamefont {Genoni}, \citenamefont {Paris},\ and\
  \citenamefont {Ferraro}}]{Albarelli2018}%
  \BibitemOpen
  \bibfield  {author} {\bibinfo {author} {\bibfnamefont {Francesco}\
  \bibnamefont {Albarelli}}, \bibinfo {author} {\bibfnamefont {Marco~G.}\
  \bibnamefont {Genoni}}, \bibinfo {author} {\bibfnamefont {Matteo G.~A.}\
  \bibnamefont {Paris}}, \ and\ \bibinfo {author} {\bibfnamefont {Alessandro}\
  \bibnamefont {Ferraro}},\ }\bibfield  {title} {\enquote {\bibinfo {title}
  {Resource theory of quantum non-gaussianity and wigner negativity},}\ }\href
  {\doibase 10.1103/physreva.98.052350} {\bibfield  {journal} {\bibinfo
  {journal} {Phys. Rev. A}\ }\textbf {\bibinfo {volume} {98}},\ \bibinfo
  {pages} {052350} (\bibinfo {year} {2018})}\BibitemShut {NoStop}%
\bibitem [{\citenamefont {Yadin}\ \emph {et~al.}(2018)\citenamefont {Yadin},
  \citenamefont {Binder}, \citenamefont {Thompson}, \citenamefont
  {Narasimhachar}, \citenamefont {Gu},\ and\ \citenamefont {Kim}}]{Yadin2018}%
  \BibitemOpen
  \bibfield  {author} {\bibinfo {author} {\bibfnamefont {Benjamin}\
  \bibnamefont {Yadin}}, \bibinfo {author} {\bibfnamefont {Felix~C.}\
  \bibnamefont {Binder}}, \bibinfo {author} {\bibfnamefont {Jayne}\
  \bibnamefont {Thompson}}, \bibinfo {author} {\bibfnamefont {Varun}\
  \bibnamefont {Narasimhachar}}, \bibinfo {author} {\bibfnamefont {Mile}\
  \bibnamefont {Gu}}, \ and\ \bibinfo {author} {\bibfnamefont {M.~S.}\
  \bibnamefont {Kim}},\ }\bibfield  {title} {\enquote {\bibinfo {title}
  {Operational resource theory of continuous-variable nonclassicality},}\
  }\href {\doibase 10.1103/physrevx.8.041038} {\bibfield  {journal} {\bibinfo
  {journal} {Phys. Rev. X}\ }\textbf {\bibinfo {volume} {8}},\ \bibinfo {pages}
  {041038} (\bibinfo {year} {2018})}\BibitemShut {NoStop}%
\bibitem [{\citenamefont {Zhuang}\ \emph {et~al.}(2018)\citenamefont {Zhuang},
  \citenamefont {Shor},\ and\ \citenamefont {Shapiro}}]{Zhuang2018}%
  \BibitemOpen
  \bibfield  {author} {\bibinfo {author} {\bibfnamefont {Quntao}\ \bibnamefont
  {Zhuang}}, \bibinfo {author} {\bibfnamefont {Peter~W.}\ \bibnamefont {Shor}},
  \ and\ \bibinfo {author} {\bibfnamefont {Jeffrey~H.}\ \bibnamefont
  {Shapiro}},\ }\bibfield  {title} {\enquote {\bibinfo {title} {Resource theory
  of non-gaussian operations},}\ }\href {\doibase 10.1103/physreva.97.052317}
  {\bibfield  {journal} {\bibinfo  {journal} {Phys. Rev. A}\ }\textbf {\bibinfo
  {volume} {97}},\ \bibinfo {pages} {052317} (\bibinfo {year}
  {2018})}\BibitemShut {NoStop}%
\bibitem [{\citenamefont {Kwon}\ \emph {et~al.}(2019)\citenamefont {Kwon},
  \citenamefont {Tan}, \citenamefont {Volkoff},\ and\ \citenamefont
  {Jeong}}]{Kwon2018}%
  \BibitemOpen
  \bibfield  {author} {\bibinfo {author} {\bibfnamefont {Hyukjoon}\
  \bibnamefont {Kwon}}, \bibinfo {author} {\bibfnamefont {Kok~Chuan}\
  \bibnamefont {Tan}}, \bibinfo {author} {\bibfnamefont {Tyler}\ \bibnamefont
  {Volkoff}}, \ and\ \bibinfo {author} {\bibfnamefont {Hyunseok}\ \bibnamefont
  {Jeong}},\ }\bibfield  {title} {\enquote {\bibinfo {title} {Nonclassicality
  as a quantifiable resource for quantum metrology},}\ }\href {\doibase
  10.1103/PhysRevLett.122.040503} {\bibfield  {journal} {\bibinfo  {journal}
  {Phys. Rev. Lett.}\ }\textbf {\bibinfo {volume} {122}},\ \bibinfo {pages}
  {040503} (\bibinfo {year} {2019})}\BibitemShut {NoStop}%
\end{thebibliography}
%

\end{document}